\documentclass{aa}  

\usepackage{graphicx}
\usepackage{txfonts}
\usepackage{natbib}
\bibpunct{(}{)}{;}{a}{}{,}
\usepackage{color}

\newcommand{\hii}{H{\sc ii}}
\newcommand{\ha}{\ifmmode {\rm H}\alpha \else H$\alpha$\fi}
\newcommand{\hb}{\ifmmode {\rm H}\beta \else H$\beta$\fi}
\newcommand{\hg}{\ifmmode {\rm H}\gamma \else H$\gamma$\fi}
\newcommand{\lya}{\ifmmode {\rm Ly}\alpha \else Ly$\alpha$\fi}

\def\fesclyc{$f_{\mathrm{esc}}^{\mathrm{LyC}}$}

\def\hii{\ion{H}{ii}}
\def\oii{[\ion{O}{ii}]}
\def\oiil{[\ion{O}{ii}]$\lambda 3727$}
\def\oiill{[\ion{O}{ii}]$\lambda\lambda 3726,3729$}
\def\oiii{[\ion{O}{iii}]}
\def\neiii{[\ion{Ne}{iii}]}
\def\neiiil{[\ion{Ne}{iii}]$\lambda 3869$}
\def\oiiil{[\ion{O}{iii}]$\lambda 5007$}
\def\oiiia{[\ion{O}{iii}]$\lambda 4363$}
\def\oiiill{[\ion{O}{iii}]$\lambda\lambda 4959,5007$}

\newcommand{\Mstar}{\ensuremath{\mathrm{M_*}}} 
 
\newcommand{\Msun}{\ensuremath{\mathrm{M}_\odot}} 
\newcommand{\Zsun}{\ensuremath{\mathrm{Z}_\odot}} 
\newcommand{\Z}{\ensuremath{\mathrm{Z}}}

\newcommand{\sSFR}{\ensuremath{\mathrm{sSFR}}}

\newcommand{\SFR}{\ensuremath{\mathrm{SFR}}}
\newcommand{\SFRs}{\ensuremath{\mathrm{SFRs}}}

\newcommand{\tauv}{\ensuremath{\mathrm{\tau}_{V}}}
\newcommand{\rtt}{R{\small 23}}
\newcommand{\ott}{O{\small 32}}
\newcommand{\Te}{\ensuremath{\mathrm{T}_{e}}}

\begin{document}

\title{Properties and redshift evolution of star-forming galaxies with high \oiii/\oii\ ratios with MUSE at $0.28 < z < 0.85$}

   \author{M. Paalvast
          \inst{1}
          \and
          A. Verhamme
          \inst{2,3}
          \and
          L.A. Straka
          \inst{1}
          \and 
          J. Brinchmann
          \inst{1}
          \and
          E.C. Herenz
          \inst{4,5}
          \and  
          D. Carton
          \inst{2}
          \and
          M.L.P. Gunawardhana
          \inst{1}
          \and          
          L.A. Boogaard
          \inst{1}
          \and    
          S. Cantalupo
          \inst{6}
          \and    
          T. Contini
          \inst{7}
           \and           
          B. Epinat
          \inst{7,8}
          \and           
          H. Inami
          \inst{2}
          \and          
          R.A. Marino
          \inst{6}
          \and           
          M.V. Maseda
          \inst{1}
          \and           
          L. Michel-Dansac
          \inst{2}
          \and           
          S. Muzahid
          \inst{1}
          \and           
          T. Nanayakkara
          \inst{1}
           \and           
          G. Pezzulli
          \inst{6}
          \and           
          J. Richard
          \inst{2}
          \and           
          J. Schaye
          \inst{1}
          \and         
          M.C. Segers
          \inst{1}
          \and           
          T. Urrutia
          \inst{4}
          \and           
          M. Wendt
          \inst{4, 9}
          \and           
          L. Wisotzki
           \inst{4}
          }

   \institute{Leiden Observatory, Leiden University, PO Box 9513, 2300 RA Leiden, The Netherlands\\
              \email{paalvast@strw.leidenuniv.nl}
             \and
                Univ Lyon, Univ Lyon1, Ens de Lyon, CNRS, Centre de Recherche Astrophysique de Lyon UMR 5574, 69230 Saint-Genis-Laval,
France 
            \and
                Observatoire de Gen\`eve, Universit\'e de Gen\`eve, 51 Ch. des Maillettes, 1290 Versoix, Switzerland
            \and 
            Leibniz-Institut f\"ur Astrophysik Potsdam (AIP), An der Sternwarte 16, 14482 Potsdam, Germany 
            \and
            Department of Astronomy, Stockholm University, AlbaNova University Centre, SE-106 91, Stockholm, Sweden
            \and
            ETH Zurich, Department of Physics, Wolfgang-Pauli-Str. 27, 8093 Zurich, Switzerland
            \and
            Institut de Recherche en Astrophysique et Plan\'etologie (IRAP), Universit\'e de Toulouse, CNRS, UPS, F-31400 Toulouse, France  
            \and
            Aix-Marseille Univ., CNRS, LAM, Laboratoire d'Astrophysique de Marseille, Marseille, France               
           \and
            Institut f\"ur Physik und Astronomie, Universit\"at Potsdam, Karl-Liebknecht-Str. 24/25, 14476 Golm, Germany
            }

   \date{Received ; accepted }
   \authorrunning{Paalvast et al.}{ }
   \titlerunning{\ott\ ratios of galaxies in MUSE data}{ }

  \abstract
  {We present a study of the \oiii/\oii\ ratios of star-forming galaxies drawn from Multi-Unit Spectroscopic Explorer (MUSE) data spanning a redshift range $0.28 < z < 0.85$. Recently discovered Lyman continuum (LyC) emitters have extremely high oxygen line ratios: \oiiil/\oiill $> 4$. Here we aim to understand the properties and the occurrences of galaxies with such high line ratios. Combining data from several MUSE Guaranteed Time Observing (GTO) programmes, we select a population of star-forming galaxies with bright emission lines, from which we draw 406 galaxies for our analysis based on their position in the $z-$ dependent star formation rate (\SFR ) - stellar mass (\Mstar ) plane. Out of this sample 15 are identified as extreme oxygen emitters based on their \oiii/\oii\ ratios (3.7$\%$) and 104 galaxies have \oiii/\oii\ > 1 (26$\%$).  Our analysis shows no significant correlation between \Mstar , \SFR,\ and the distance from the \SFR\ - \Mstar\ relation with \oiii/\oii . We find a decrease in the fraction of galaxies with \oiii/\oii\ > 1 with increasing \Mstar , however, this is most likely a result of the relationship between \oiii/\oii\ and metallicity, rather than between  \oiii/\oii\ and \Mstar . We draw a comparison sample of local analogues with $<z> \approx 0.03$ from the Sloan Digital Sky Survey, and find similar incidence rates for this sample. In order to investigate the evolution in the fraction of high \oiii/\oii\ emitters with redshift, we bin the sample into three redshift subsamples of equal number, but find no evidence for a dependence on redshift. Furthermore, we compare the observed line ratios with those predicted by nebular models with no LyC escape and find that most of the extreme oxygen emitters can be reproduced by low metallicity models. The remaining galaxies are likely LyC emitter candidates. Finally, based on a comparison between electron temperature estimates from the \oiiia/\oiiil\ ratio of the extreme oxygen emitters and nebular models, we argue that the galaxies with the most extreme \oiii/\oii\ ratios have young light-weighted ages.   
 }


   \maketitle

\section{Introduction}
\label{s_intro}

Extreme oxygen line ratios (\ott\ $\equiv$ \oiiil/\oiil $>4$) were proposed recently as a potential tracer of the escape of ionising radiation from galaxies through density-bounded \hii\ regions \citep{Jaskot13, Nakajima14}. The idea is the following: if a galaxy is leaking ionising photons through a density-bounded region, the ratio of \ott\ can be high if the \hii\ region that we observe is truncated, for example if (part of) the \oii\ region is missing. We see deeper into the ionised region and the external layer of \oii\  is either nonexistent or thinner than in the classical ionisation-bounded scenario. Given their high \ott\ ratios, \citet{Jaskot13} discuss the possibility of LyC escape from "Green Pea" (GP) galaxies, a population of extremely compact, strongly star-forming galaxies in the local Universe \citep{Cardamone09, Izotov11, 2016ApJ...820..130Y}. \citet{Nakajima14} and \citet{Nakajima16} compare \ott\ ratios of different types of high-redshift galaxies, Lyman Break Galaxies (LBGs), and Lyman Alpha Emitters (LAEs) with GPs and Sloan Digital Sky Survey (SDSS) galaxies: high-redshift galaxies have on average higher \ott\ ratios than SDSS galaxies, but comparable to GPs. Furthermore, the observed \ott\ ratios of LAEs are larger than those of LBGs. Along the same line, GPs are also strong LAEs \citep{Henry15, 2016ApJ...820..130Y, 2017A&A...597A..13V, 2017ApJ...838....4Y}, which is very unusual for galaxies in the local Universe \citep{Hayes11, Wold14}. 
 
While the \ott\ ratio of galaxies that are leaking ionising photons may be enhanced compared to those with a LyC escape fraction, \fesclyc , equal to zero, there are other situations that can lead to high \ott\ ratios. For example, the  \ott\ ratio depends on metallicity: low stellar and nebular metallicities lead to higher \ott\ ratios \citep{Jaskot13}. A harder ionising spectrum will also induce higher \ott\ ratios, as investigated in for example \citet{Pellegrini12}, as well as a higher ionisation parameter (e.g. \citealt{Stasinska15}). Furthermore, shocks could also explain these ratios, as studied in detail in \citet{Stasinska15}.

Despite intensive searches for LyC emission from galaxies, only a few LyC leakers have been identified over the last decades in the local Universe \citep{Bergvall06, Leitet13, Borthakur14, Leitherer16}, but most searches resulted in non-detections or upper limits \citep{Siana15, Mostardi15, Grazian16, Rutkowski16, Rutkowski17}. The discovery of the link between LyC emission and \ott,\ however, turned the tide, as demonstrated by for example \citet{Izotov16a,Izotov16b, 2018MNRAS.474.4514I, 2018MNRAS.tmp.1318I}. For their studies, LyC emission was detected for all eleven galaxies at $z \approx 0.3$ that were selected by their extreme \ott\ ratios (\ott\ > 4), among other criteria such as brightness, compactness, and strong \hb\ equivalent widths. Furthermore, a correlation between \ott\ and the escape of ionising photons was found, although the scatter of \fesclyc\ is large \citep{2018MNRAS.tmp.1318I}. At high redshift ($z \approx 3$), four galaxies with high escape fractions ($> 50$\%)  have been reported \citep{Vanzella15, 2016A&A...585A..51D, Shapley16, Bian17, 2018MNRAS.476L..15V}, which were selected by similar criteria. Additionally, the recent results from the Lyman Continuum Escape Survey \citep{2018arXiv180601741F} reveal an average escape fraction of $\sim20\%$ for galaxies at $z\approx3$ with strong \oiii\ emission, and a weak correlation between \fesclyc and the \oiii\ equivalent width for $\sim20$ galaxies with directly detected LyC emission. Although the combination of these selection criteria has resulted in relatively few galaxies with confirmed LyC emission yet, the detection of extreme \ott\ emission from a local low-mass GP analogue \citep{2017ApJ...845..165M}  might, however, suggest that low-mass extreme \ott\ emitters, and thus possible low-mass LyC emitters, are more common than the bright GP samples suggest. A statistical study of the \ott\ ratios of emission-line selected galaxies over a broad range of stellar masses has, however, not been performed so far. 

The unique capabilities of the Multi-Unit Spectroscopic Explorer (MUSE) \citep{2010SPIE.7735E..08B} allow us to study the properties of galaxies with extreme \ott\ ratios and how common they are in emission-line selected samples.  For this study we combine four MUSE Guaranteed Time Observing (GTO) surveys and collect a sample of mainly emission-line detected galaxies with a high specific star formation rate and stellar masses between $\sim10^6$ and $\sim10^{10}$, from which we compute the distribution of \ott\ ratios in a blind survey of star-forming galaxies. We will here present the properties and occurrences of extreme oxygen emitters spanning the redshift range 0.28 < z < 0.85, where both lines are in the MUSE spectral range, in the largest statistical sample of emission-line detected galaxies in three-dimensional spectral data.

This article is organised as follows: in Sect.~\ref{s_data} we describe the data from different programmes that we used for this study; in Sect.~\ref{s_sample} we describe the sample selection; in Sect.~\ref{s_results} we investigate the occurrence of high \ott\ ratios and study potential correlations with stellar mass ( \Mstar ), star formation rate (\SFR ), and the metallicity indicator \rtt\ line ratio; in Sect.~\ref{s_discussion} we study the incidence rate of galaxies with high \ott\ ratios as a function of \Mstar\ and $z$, and we also discuss how our results compare to nebular models with no escape of ionising photons. We end with a discussion on the most extreme oxygen emitters. Throughout this paper we adopt a cosmology with $H_0$ = 70 km s$^{-1}$ Mpc$^{-1}$, $\Omega_m$ = 0.3 and $\Omega_\Lambda$ = 0.7. 

\section{Data}
\label{s_data}
\subsection{Different MUSE surveys}
For this study we used data taken with MUSE,  as part of GTO observations, covering a wavelength range of 4800-9300 $\AA$. We selected data from four surveys that together span an area of more than 55 arcmin$^2$. Below follows a description of the different surveys. 

\subsubsection{Hubble Ultra Deep Field survey}
The  MUSE Hubble Ultra Deep (HUDF) survey \citep{2017A&A...608A...1B} consists of two fields of different size and depth in the original HUDF region. The Medium Deep Mosaic Field (e.g. UDF-mosaic) is a mosaic of nine pointings ($\approx$3$\arcmin$ x 3$\arcmin$) with a depth of approximately ten hours. The UDF-10 Ultra Deep Field (e.g. UDF-10) is a deeper observation of a single pointing within the UDF-mosaic region, with a depth of approximately 31 hours, covering $\approx$1.15 arcmin$^2$. The data reduction is described by \citet{2017A&A...608A...1B} and is based on the MUSE standard pipeline version 1.7dev \citep{2014ASPC..485..451W}. For the construction of the redshift catalogue \citep{2017A&A...608A...2I}, Hubble Space Telescope (HST) priors are used, as well as a blind search for emission lines in the datacube using the software ORIGIN (Mary et al. in prep).

\subsubsection{MUSE-Wide survey}
The MUSE-Wide survey aims to cover a larger field of view than the HUDF survey with relatively short exposures of one hour per pointing. The data release of the full survey will be presented in Urrutia et al (in prep). Here we focus on the first 24 pointings of MUSE-Wide, which together cover an area of 22.2 arcmin$^2$. We use the source catalogue that is presented in \citet{2017A&A...606A..12H}, which is created using the emission-line detection software LSDCat \citep{2017A&A...602A.111H}, but we do not use their supplied spectra since we extract the spectra of all surveys consistently. 

\subsubsection{MUSE QuBES survey}
The MUSE Quasar Blind Emitter Survey (QuBES) consists of 24 individual fields centred on quasars. All datacubes have a minimum total exposure time of two hours, with selected fields observed to a depth of ten hours based on the availability of high-quality archival auxiliary data such as quasi-stellar object (QSO) spectra. The standard data reduction was carried out with the MUSE Data Reduction System (DRS; \citealt{2014ASPC..485..451W}). Post-processing procedures for additional integral field unit (IFU) normalisation and sky subtraction were carried out using the CubEx package (Cantalupo et al. in prep; see \citealt{2018ApJ...859...53M} and \citealt{2016ApJ...831...39B} for details). The survey will be fully described in Straka et al. (in prep) and Segers et al. (in prep). A subset of 21 fields is used for this study based on the availability of galaxy catalogues. For this study, the presence of the QSO in the field is not important. The galaxy catalogues are compiled as follows. First, white light images are created from the MUSE datacubes by summing the entire cube along the wavelength axis. Then, SExtractor \citep{1996A&AS..117..393B} is used to detect any objects down to 1$\sigma$. The spectra for each object are extracted by selecting associated pixels in the segmentation maps produced by SExtractor. The spectra of the resulting detections are then inspected by eye in order to determine the galaxy redshifts based on nebular emission lines and stellar absorption features. 

\subsubsection{Galaxy groups survey}
The last dataset added to our sample is that of the Galaxy group survey (Epinat et al. in prep), which targets galaxy groups at intermediate redshift ($z \approx 0.5-0.7$) from the zCOSMOS 20k group catalogue \citep{2012ApJ...753..121K}. We selected data from three galaxy groups, namely COSMOS-Gr30, 34, and 84, with the deepest MUSE data of 9.75, 5.25, and 5.25 hours respectively, \textasciitilde 1$\arcmin$ x 1$\arcmin$ each. The data reduction followed the same approach as the Hubble Ultra Deep Field and is described in \citet{2018A&A...609A..40E} for the galaxy group field COSMOS-Gr30. For the construction of the redshift catalogues, galaxies were selected from the COSMOS photometric catalogue by \citet{2016ApJS..224...24L}, complemented by emission-line detection using ORIGIN for the deepest field COSMOS-Gr30 (see also \citealt{2018A&A...609A..40E}). 

\subsection{Spectrum extraction and emission-line flux measurements}
The spectra of all sources are extracted from the datacubes using the same method in order to make the line flux measurements comparable. We followed the approach of \citet{2017A&A...608A...2I} and extracted the spectra using a mask region, which is the HST segmentation map convolved by the MUSE point spread function for all surveys except the MUSE QuBES survey, for which there is no HST coverage. Spectra in this survey are extracted using a mask region that is constructed from the MUSE white light image. We then used the simple unweighted sum of the flux in the mask region and used this as the spectrum of each galaxy. We measured the line fluxes of the galaxies in the catalogues using  the software \textsc{platefit} \citep{2004ApJ...613..898T, 2008A&A...485..657B}. Since this method is the same as that used by \citet{2017A&A...608A...2I} to construct the HUDF emission-line catalogue, the line flux measurements that we use here are identical to theirs. 

\subsection{Deriving \SFRs\ and dust extinction}
The \SFRs\ of our galaxies are calculated using the method described in \citet{2013MNRAS.432.2112B}. In short, we simultaneously fit the \citet{2001MNRAS.323..887C} models to the brightest (signal-to-noise ratio (S/N)$>$3) optical emission lines. From this we estimate the \SFR\ marginalising over: metallicity, ionisation parameter, dust-to-metal ratio, and the optical depth of the dust attenuation (\tauv). The advantage of using a multi-emission-line approach over using a single Balmer line to calculate the \SFR\ is that it is less affected by sky line contamination of a single line and should therefore provide a more robust \SFR . Also, this method provides an estimate of \tauv , which we adopt to correct the emission-line fluxes for dust extinction. 

\subsection{Calculating stellar masses}
We obtained stellar masses for the galaxies by performing spectral energy distribution (SED) fitting using the \textsc{fast} (Fitting and Assessment of Synthetic Templates) algorithm \citep{2009ApJ...700..221K}, where we used the \citet{2003MNRAS.344.1000B} library and assumed exponentially declining star formation histories (SFHs) with a \citet{2000ApJ...533..682C} extinction law and a \citet{2003ApJ...586L.133C} initial mass function (IMF). To test the influence of our SFH assumption, we compare our stellar masses with those derived by the \textsc{magphys} code \citep{2008MNRAS.388.1595D}. For these \Mstar\ estimations, the photometry is fitted to stellar population synthesis models assuming random bursts of star formation in addition to an exponentially declining SFH. We find consistent stellar masses, for example the median difference between \Mstar\ derived from \textsc{fast} and \Mstar\ from \textsc{magphys} equals $\log \Mstar/\Msun = 0.08$ with a standard deviation of $\log \Mstar/\Msun = 0.3$,  from which we conclude that the influence of the assumed SFH is small for this sample. For the UDF, we used HST Advanced Camera for Surveys (ACS) and  Wide Field Camera 3 (WFC3) photometry from the catalogue of \citet{2015AJ....150...31R}. The same approach is applied to the MUSE-Wide (photometry from \citealt{2014ApJS..214...24S}) and the Galaxy groups (photometry from \citealt{2016ApJS..224...24L}). Unfortunately there is no deep photometry available for the MUSE-Qubes survey. We therefore used a set of 11 400 $\AA$ -wide boxcar filters, where emission lines are masked, to compute a pseudo-photometric SED, as will be described in more detail in Segers et al. (in prep). Since not all the bright emission lines lie within the MUSE spectral range for our redshifts, we decided to leave the photometry uncorrected for bright emission lines. To test the possible effect of strong emission lines on \Mstar , we compared our stellar masses to those based on the photometry that is corrected for emission lines in the MUSE spectral range.  We find that for the bulk of the galaxies in our sample, this effect is negligible; for example, the maximum difference between the emission-line corrected and the non-corrected \Mstar\  for the extreme \ott\ emitters, which will be introduced in Sect. \ref{s_results}, corresponds to $\log$ \Mstar/\Msun = 0.03.

\subsection{Redshift distribution}
\begin{figure*}
 \centering
    \includegraphics[width=\hsize]{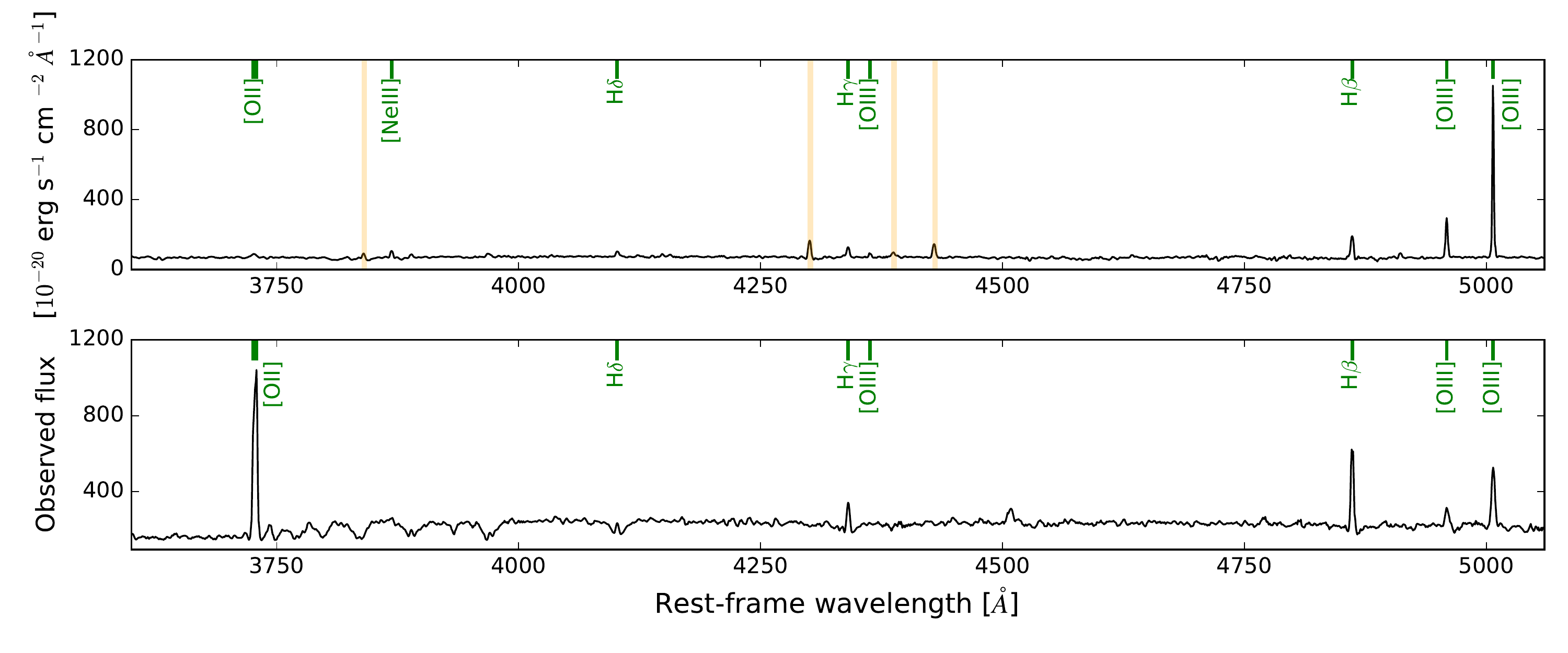}
      \caption{Example spectra of \ott\ emitters in our sample. The upper panel shows the spectrum of the galaxy with the most extreme oxygen ratio (dust-corrected \ott\ = 23) from the UDF-mosaic catalogue with id = 6865 and z = 0.83. The orange shaded lines show the \hg , \hb,\ and \oiiill\ of neighbouring sources at $z\approx0.62$. An example spectrum of a galaxy with a lower \ott\ ratio, in this case with dust-corrected \ott\ = 0.25, is shown in the lower panel (UDF-mosaic, id = 892, z = 0.74).}
      \label{fig:example_spec}
\end{figure*}

\begin{figure}
 \centering
   \includegraphics[width=\hsize]{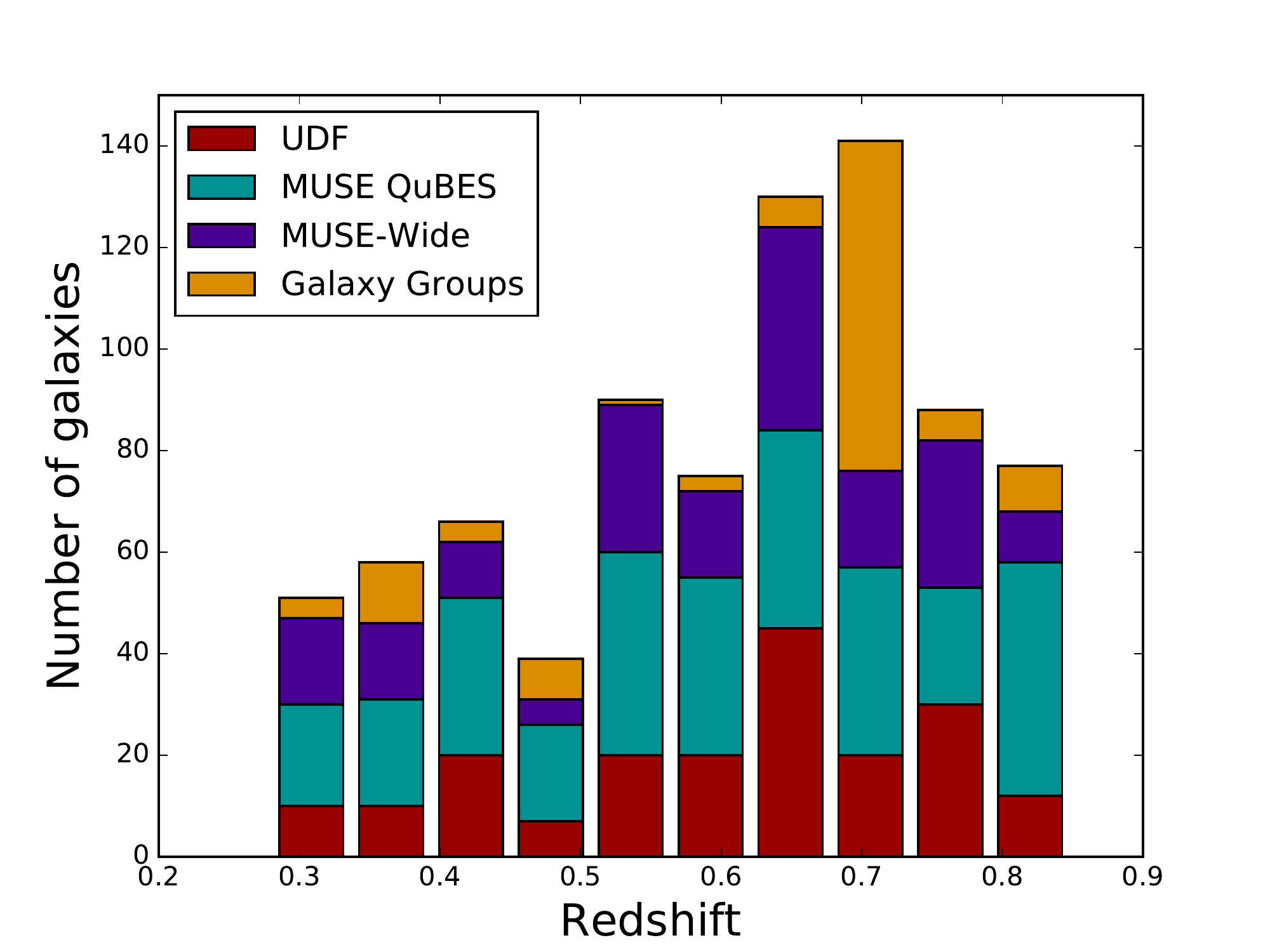}
      \caption{Redshift distribution of 815 galaxies in total, in the range 0.28 < z < 0.85, with high confidence, spectroscopically determined MUSE redshift, in the four surveys that we are using for this project.}
      \label{fig:galaxies_histogram}
\end{figure}

For our main sample, we only select galaxies with spectra of sufficient quality for our study, that is, spectra with S/N > 3 in the \oiiil\ line at $0.28 < z < 0.85,$ since for this redshift interval the \oii\ and \oiii\ emission lines fall within the MUSE wavelength range. For galaxies that meet this criterion, but have S/N(\oii ) < 3, we use the 3-$\sigma$ lower limit for the \ott\ ratio. We show two examples of spectra that meet these criteria in Fig. \ref{fig:example_spec}. This leads to a total sample of 815 galaxies, of which the redshift distribution for each survey is shown in Fig. \ref{fig:galaxies_histogram}. This histogram shows the number of galaxies in the redshift bins, separated based on the survey from which the data originates. Because the depths of the different surveys that we combine here are not uniform, care is required for selecting galaxies for our final sample. 

\subsection{Sample selection}
\label{s_sample}
\label{sec:selection}
\begin{figure*}
 \centering
   \includegraphics[width=\textwidth]{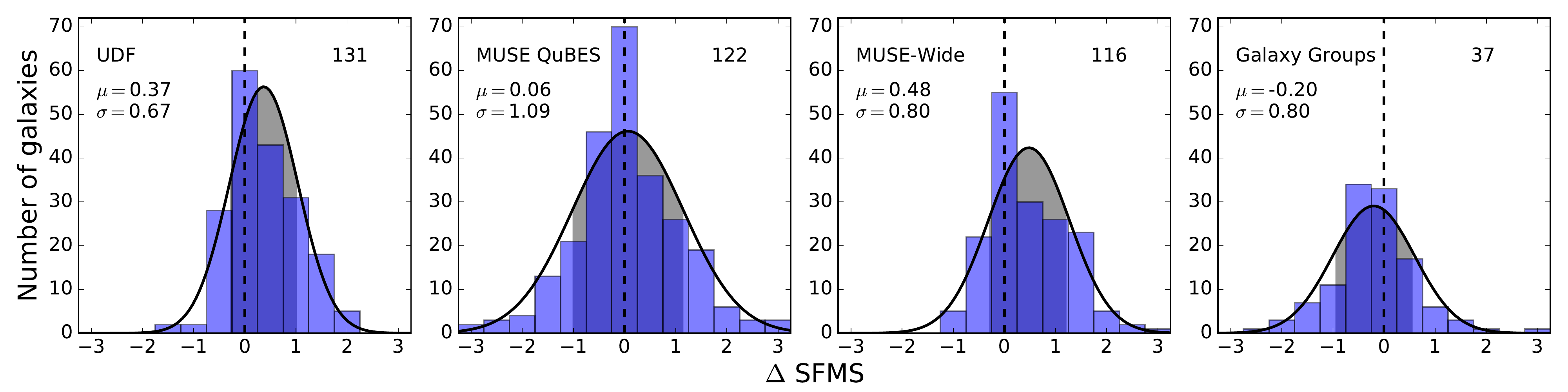}
      \caption{Distribution of the distance to the redshift-dependent SFMS from \citet{Boogaardetal} (Equation \ref{SFMS}) in dex of the four different surveys used for this study. We show the 1-$\sigma$ variation around the mean $\mu$ (grey area) calculated from the fitted normal distribution (black line).  We included all galaxies with a distance > 0 dex (dashed line), so all galaxies that lie above the main sequence. The number in the upper right corner of the diagram shows the number of galaxies above this threshold. In our final sample we include 406 galaxies in total.}   
      \label{fig:distr_ssfr}
\end{figure*}

Observations of star-forming galaxies have shown that the star formation rate (\SFR) and stellar mass (\Mstar) of these galaxies are tightly related (e.g. \citealt{2004MNRAS.351.1151B, 2007ApJ...660L..43N}). This relation, also called the star-forming main sequence (SFMS), has been studied intensively in the last decade (e.g. \citealt{2012ApJ...754L..29W, 2014ApJ...795..104W, 2015ApJ...801...80L, 2017ApJ...847...76S}). \citet{Boogaardetal} have constrained the SFMS for galaxies that are detected in deep MUSE data, modelling them as a Gaussian distribution around a three-dimensional (3-D) plane, taking into account and obtaining the redshift evolution of the \SFR\ - \Mstar\ relation, resulting in
\begin{equation}
\label{SFMS}
\begin{aligned}
\log \SFR\ = {} & 0.83^{+0.07}_{-0.06} \log \left(\frac{M_*}{M_0}\right) -0.83^{+0.05}_{-0.05} \\ 
                        & + 1.74^{+0.66}_{-0.68} \log \left(\frac{1 + z}{1 + z_0}\right) \pm 0.44^{+0.05}_{-0.04} 
\end{aligned}
,\end{equation} 
with $M_0 = 10^{8.5}$ \Mstar\ and $z_0 = 0.55$. Because this relation is derived for a sample of galaxies with deep photometry and high S/N Balmer lines from MUSE spectra of galaxies with stellar masses down to log \Mstar /\Msun\ $\approx$ 7, and comparable to the mass range of the galaxies in our sample, we use this $z$-dependent \Mstar - \SFR\ relation to select galaxies for our final sample. We calculate how much the \SFR\ of a galaxy is offset from the redshift-corrected SFMS, which we will herein refer to as the 'distance to the SFMS'  ($\Delta$ SFMS) given in dex with a sign such that objects above the SFMS have a positive distance. The distribution of the distances to the SFMS is shown in Fig. \ref{fig:distr_ssfr} for each survey that we use for this study separately. The dashed line represents the SFMS, with, at the left side, galaxies below and, at the right side, galaxies above the SFMS. For all four surveys, this distribution peaks within 1-$\sigma$ from the SFMS. Since our sample is mostly emission-line selected, we expect our sample to be most complete at high \SFRs . However, below the SFMS it is likely that a fraction of the galaxies are below the detection threshold. We therefore conservatively select galaxies with a distance > 0 dex from the SFMS, resulting in a final sample of 406 galaxies. Applying such a selection based on a fixed distance from the $z-$dependent SFMS relation also ensures that we select the same fraction of star-forming galaxies at each redshift.
For an overview of the stellar mass and the \SFR\ distribution of the four surveys, we refer the reader to Figs. \ref{fig:app_distr_m} and \ref{fig:app_distr_sfr} in the Appendix. 

Based on the stellar mass distribution, we assume that all the galaxies in the sample are pure star-forming systems, since studies show that the fraction of active galactic nucleus (AGN) host galaxies is low at these masses. For example almost all AGN hosts at $0 < z < 1$ have $\log$ \Mstar/\Msun\ $\gtrsim 10.2$ \citep{2013A&A...556A..11V} and the fraction of AGNs over all galaxies with $\log$ \Mstar/\Msun\ $\approx 10$ at $z \lesssim 0.3 $ is $\sim$1 $\%$ \citep{2003MNRAS.346.1055K,2010ApJ...723.1447H}.

\section{Results}
\label{s_results}
In this section we explore whether there is a correlation between galactic properties and \ott .  We also study the location of these emitters with respect to the SFMS, since \citet{2017A&A...605A..67C} report that LyC leakers lie further above the SFMS.
 
\subsection{Redshift distribution of extreme \ott\ emitters}
\begin{figure}
 \centering
   \includegraphics[width=\hsize]{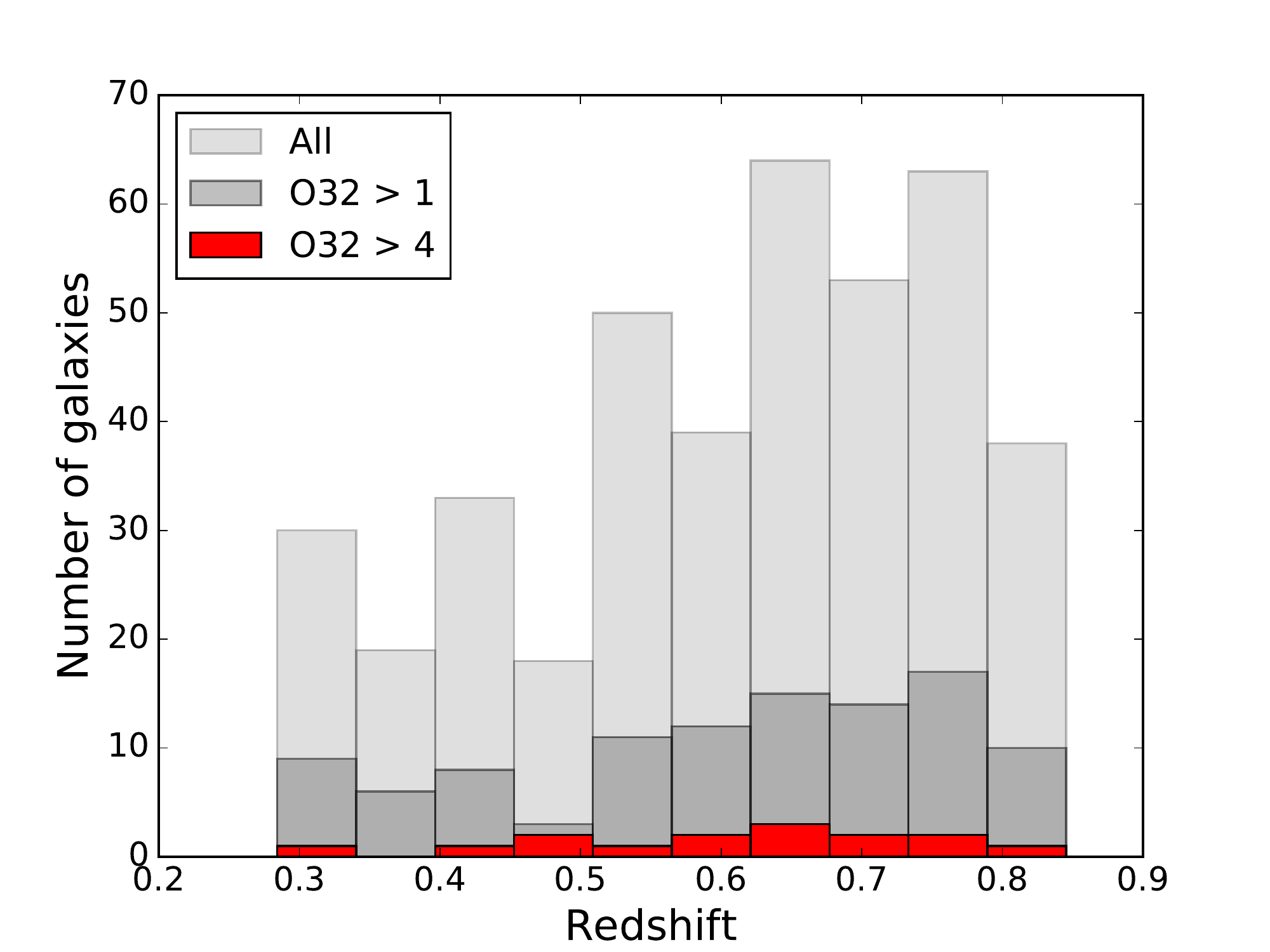}
      \caption{Redshift frequency of extreme \ott\ emitters with \ott\ > 4 (red). In dark grey we show the redshift distribution of the galaxies with  \ott\ > 1 and in light grey the total sample.}
      \label{fig:redshift_frequency}
\end{figure}

In Fig. \ref{fig:redshift_frequency} we show the redshift distribution of our sample after the selection that is described in the previous section. In light grey we show the number density of all the galaxies, while in dark grey we show that of galaxies with  \ott\ > 1. For galaxies with S/N(\oii ) $<$ 3, we use the 3-$\sigma$ upper limits on \oii , resulting in 3-$\sigma$ lower limits on \ott . We determined the ratio between the number of galaxies with \ott\ > 1 to the total number of galaxies in a bin.  A $\chi^2$ statistical test against the null hypothesis that this fraction in each bin is independent of redshift gives $\chi^2$ = 1.9 a $P$-value of $\sim$0.99, which indicates that the fraction of galaxies with oxygen ratios above unity does not evolve as a function of redshift. We adopted the same approach for the extreme oxygen emitters with \ott\ > 4, as shown in red, which yielded a comparable result ( $\chi^2$ = 4.4, $P$-value $\approx$ 0.88). This latter result is not particularly robust, because we only have 15 extreme emitters in our sample. 

\subsection{\ott\ emitters on the star-formation main sequence}
\begin{figure}
 \centering
   \includegraphics[width=\hsize]{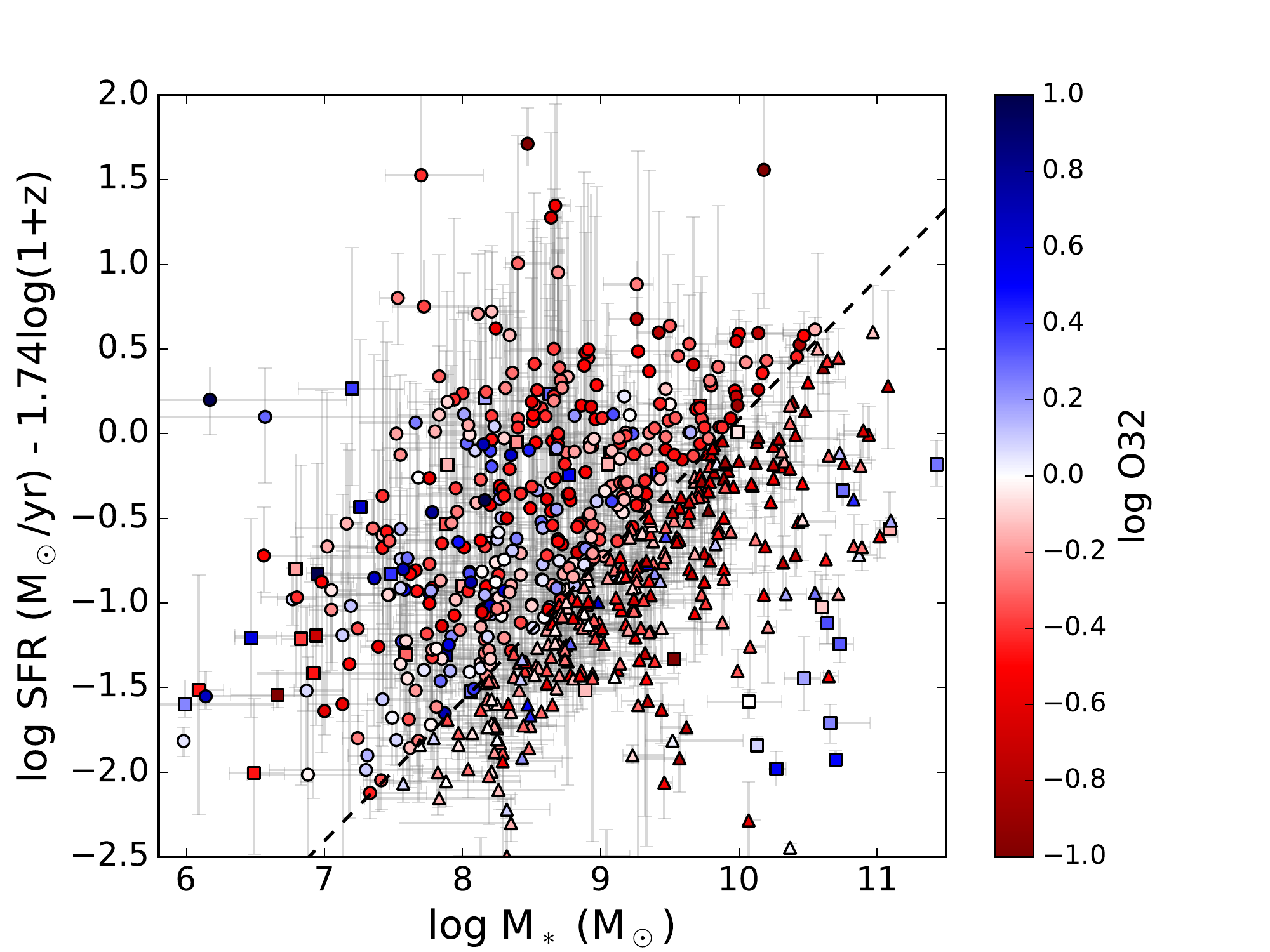}
      \caption{ \SFR\ - \Mstar\ diagram where we have corrected the \SFRs\ to $z=0$ using the redshift evolution from Eq. \ref{SFMS}. The dashed line shows this relation for $z = 0$, above which we selected galaxies in our sample, as described in Sect. \ref{sec:selection}. The points are coloured by the logarithm of the dust-corrected \ott\ ratio. The symbols show galaxies in the final selection (circles), galaxies below the selection threshold (triangles), and galaxies with a lower limit on the \ott\ ratio (squares).}
      \label{fig:main_sequence}
\end{figure}

\begin{figure}
 \centering
   \includegraphics[width=\hsize]{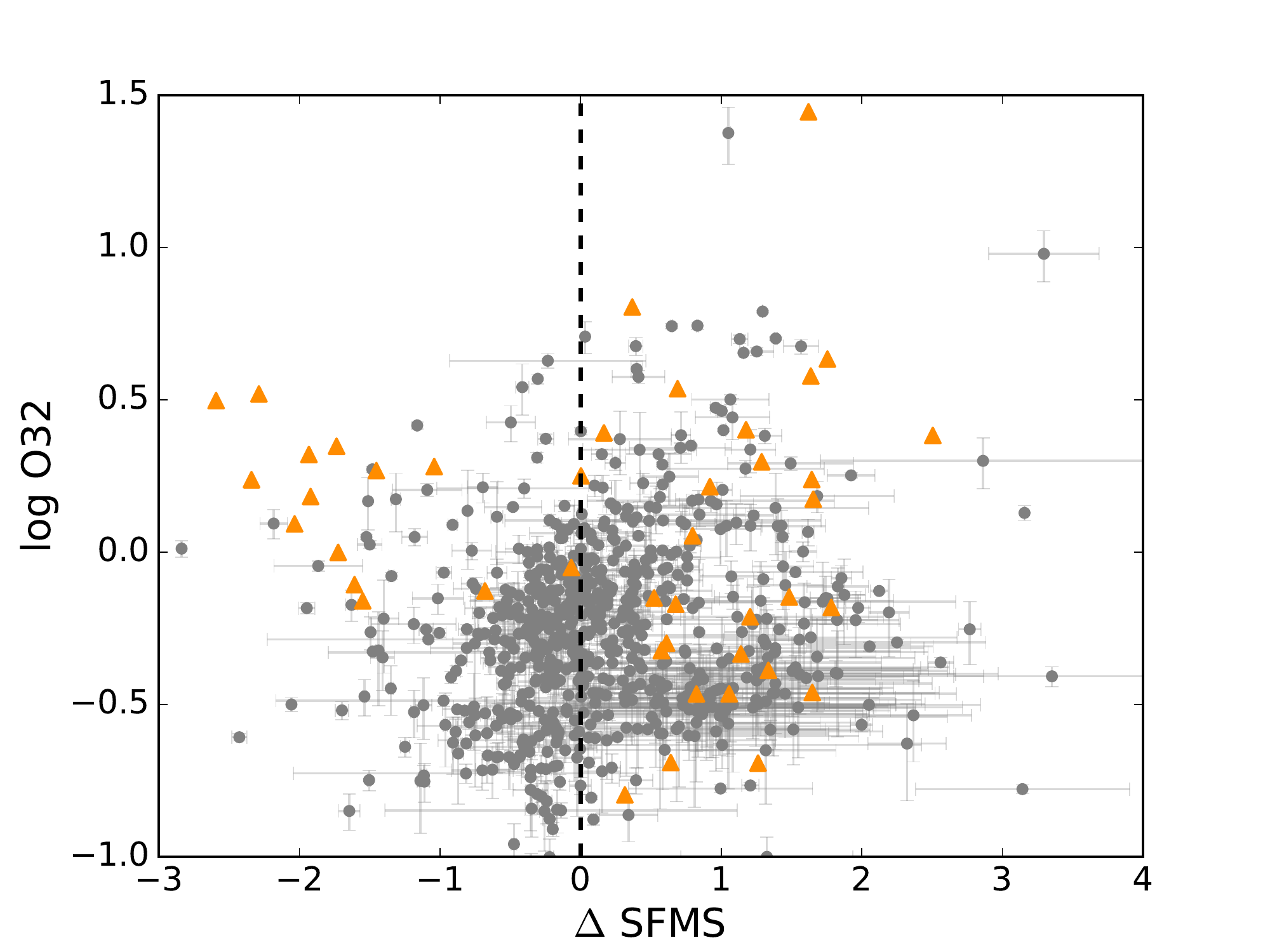}
      \caption{\ott\ ratio versus the distance to the SFMS. Only galaxies above $\Delta$ SFMS = 0 (dashed line) are included in the final sample. Lower limits on \ott\ are shown by orange triangles.}   
      \label{fig:deltasfms_o32}
\end{figure}

The redshift-corrected distribution in the $\log$ \SFR\ - $\log$ \Mstar\ plane is shown in Fig. \ref{fig:main_sequence}. We normalised our results to $z=0$, and show the SFMS of Eq. \ref{SFMS} corrected to $z=0$ (dashed line). The same diagram coloured by survey is shown in Fig. \ref{fig:app_SFMS} in the Appendix. Only galaxies above this relation are selected for the final sample (circles), but for comparison we also show the galaxies that are left out by the SFMS selection (triangles). Galaxies with oxygen ratios larger than unity (blue points) are overall more abundant above the SFMS (104 out of 406 galaxies, 26$\%$) than below this relation (46 out of 324 galaxies, 14$\%$). The number of extreme \ott\ emitters deviates even more;  15 versus one galaxies in \ott\ regime above and below the SFMS, respectively. Regarding only galaxies in the final sample, there is no clear correlation between distance from the SFMS and \ott\ ratio. Moreover, we also find that galaxies with \ott\ > 1 are more common at low masses ($\log$\Mstar/\Msun\ < 9). A two-dimensional plot of \ott\ versus the distance to the SFMS is shown in Fig. \ref{fig:deltasfms_o32}. We will now turn to a more quantitative discussion of the relation between the oxygen ratio and \SFR\ and \Mstar .

\subsection{\ott\ as a function of stellar mass}
\label{subsec:o32M}
\begin{figure}
 \centering
   \includegraphics[width=\hsize]{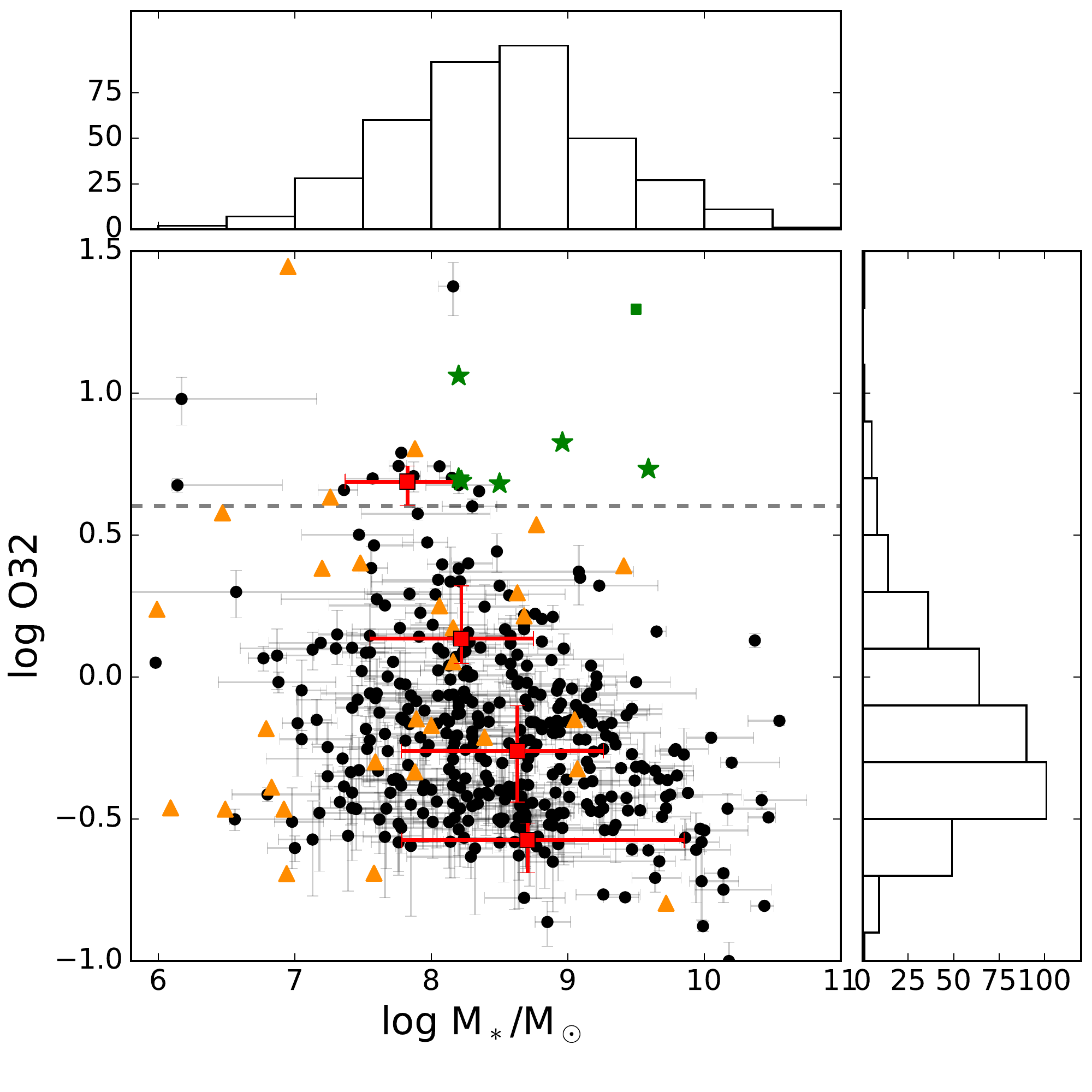}
      \caption{Stellar mass as a function of $\log$ \ott\ (black dots) and lower limits on \ott\ (orange triangles). The median per bin with 1-$\sigma$ errors is shown by the red squares and error bars. We defined the bins by intervals of  $\log$\ott\ between -1.0 and 1.0 with a step size of 0.5. Three-$\sigma$ lower limits on the \ott\ ratio for sources with a detection of \oii\ below our S/N cut are shown by the orange triangles. The green stars indicate the positions of the confirmed Lyman continuum leakers from \citet{Izotov16a,Izotov16b, 2018MNRAS.474.4514I} and the green square is a LyC emitter at $z$=3.2 \citep{2016A&A...585A..51D, 2016ApJ...825...41V}. Galaxies above the grey dashed line have extreme \ott\ ratios (\ott\ > 4). }
      \label{fig:m_o32}
\end{figure}

In Fig. \ref{fig:m_o32} we plot the $\log$ \ott\ line ratio versus the stellar mass (black dots). The red squares denote the median values in both the x- and y-directions, in the intervals of  $\log$ \ott\ between -1.0 and 1.0 with a step size of 0.5. Additionally we show 3-$\sigma$ lower limits of the \ott\ ratio for galaxies without an \oii\ detection above our S/N threshold (orange triangles). The confirmed LyC-leaking galaxies from \citet{Izotov16a,Izotov16b, 2018MNRAS.474.4514I} are shown by green stars, and the extreme LyC emitter at $z=3.2$ from \citet{2016A&A...585A..51D} and \citet{2016ApJ...825...41V} is shown by the green square. The histograms along the y- and x-axis represent the distribution of \ott\ and \Mstar , respectively. 

The median values in the \ott\ bins show a clear anti-correlation between the oxygen ratio and the stellar mass. However, the Spearman's rank correlation coefficient for individual galaxies equals -0.30 ($P$-value $\approx$ 0), indicating no clear trend between $\log$ \ott\ and  $\log$ \Mstar . The stellar masses of galaxies with \ott\ > 4 are lower than the average stellar mass of the entire population, for example all the extreme oxygen emitters in our sample have stellar masses below $10^9$. Although the \ott\ ratios of the extreme emitters in our sample are similar to those of the confirmed LyC leakers, their stellar masses are smaller than those of most of the leaking galaxies from \citet{Izotov16a, Izotov16b, 2018MNRAS.474.4514I}, \citet{2016A&A...585A..51D} and \citet{2016ApJ...825...41V}. Because their galaxies were, besides their extreme \ott\ ratios, selected by their brightness to increase the possibility to directly detect LyC photons at $z \approx 0.3$, their mass is not necessarily a reflection of the typical mass of an LyC emitter. For example, \citet{2017MNRAS.471..548I} show that galaxies at $z<0.1$, which are only selected by extreme \ott\ ratios, all have masses between $10^{6} - 10^{7} \Msun$. In addition, Mrk 71, a near green pea analogue and a LyC emitter candidate, has a stellar mass around $10^{5}$ \citep{2017ApJ...845..165M}, suggesting that the mass of the bulk of LyC emitters might  be lower than what is derived from confirmed LyC leakers. The approximately 20 recently discovered galaxies with LyC emission at $z\approx3$ \citep{2018arXiv180601741F} show an anti-correlation between the LyC escape fraction and the stellar mass, similar to our results.

\subsection{\ott\ as a function of \SFR}
\begin{figure}
 \centering
   \includegraphics[width=\hsize]{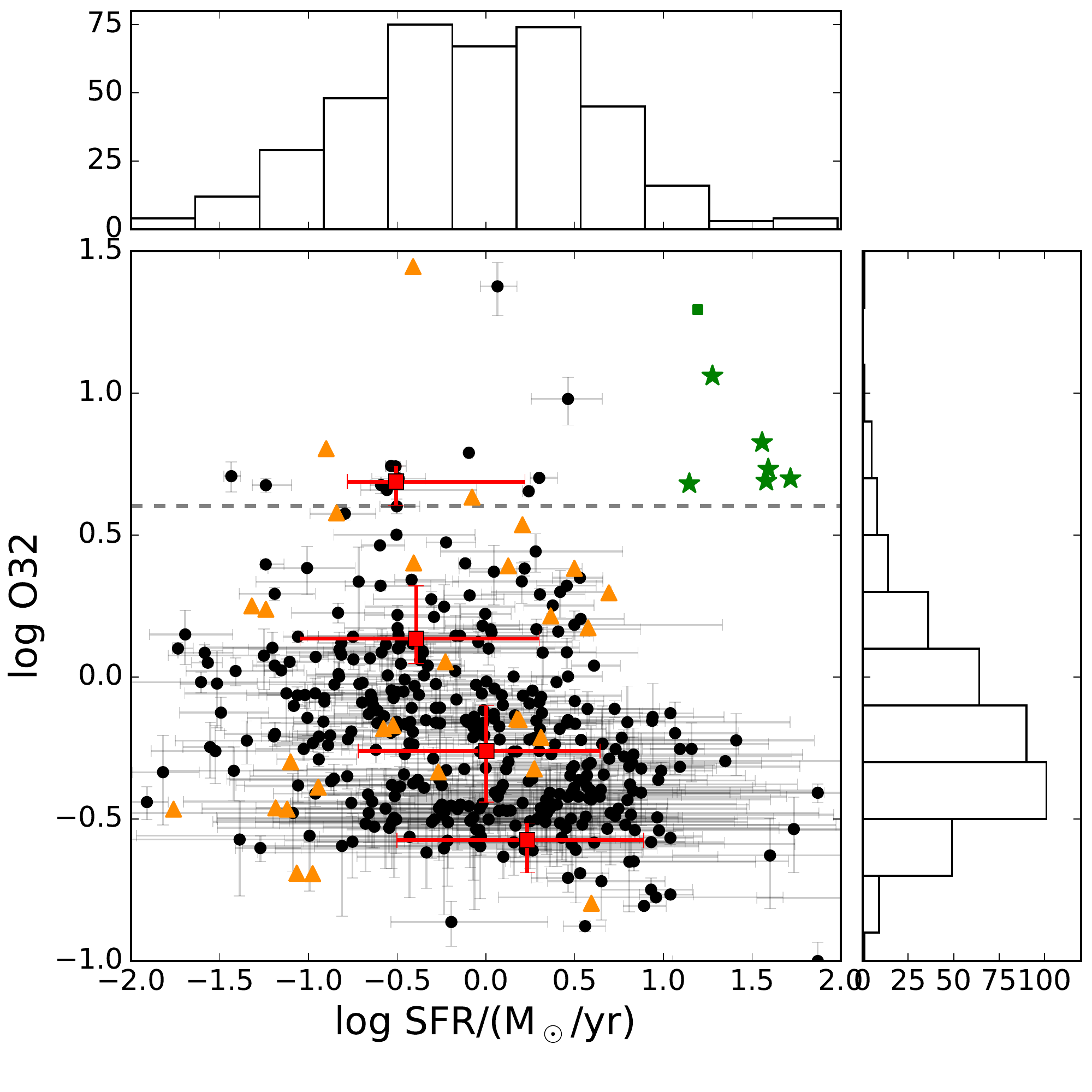}
      \caption{\ott\ ratio versus \SFR . The colours are described in the caption of Fig. \ref{fig:m_o32}. All \SFR s that are shown here are derived from emission lines, as described in  Sect. \ref{sec:selection}.}
      \label{fig:sfr_o32}
\end{figure}

In Fig. \ref{fig:sfr_o32} we show the oxygen ratio as a function of \SFR\ with the same colour code as used in Fig. \ref{fig:m_o32}. The median values of the \SFR\ decrease with increasing \ott . For individual galaxies there is again no clear correlation (Spearman's rank correlation coefficient $\approx$ -0.35, $P$-value $\approx$ 0). The \SFR\ of the confirmed LyC emitters, visualised by the green stars and square, is about two orders of magnitude larger than the median of our galaxies with comparable \ott\ emission.

When comparing the \ott\ ratio with the specific star formation rate (\sSFR\ = \SFR/\Mstar), we find that these values are also not correlated and that this also holds for the median values in the \ott\ bins. The \sSFR\ of the confirmed LyC emitters is on average one order of magnitude larger than the \sSFR\ of our sample. 

\subsection{\ott\ as a function of \rtt}
\label{results:r23}
\begin{figure}
   \centering
   \includegraphics[width=\hsize]{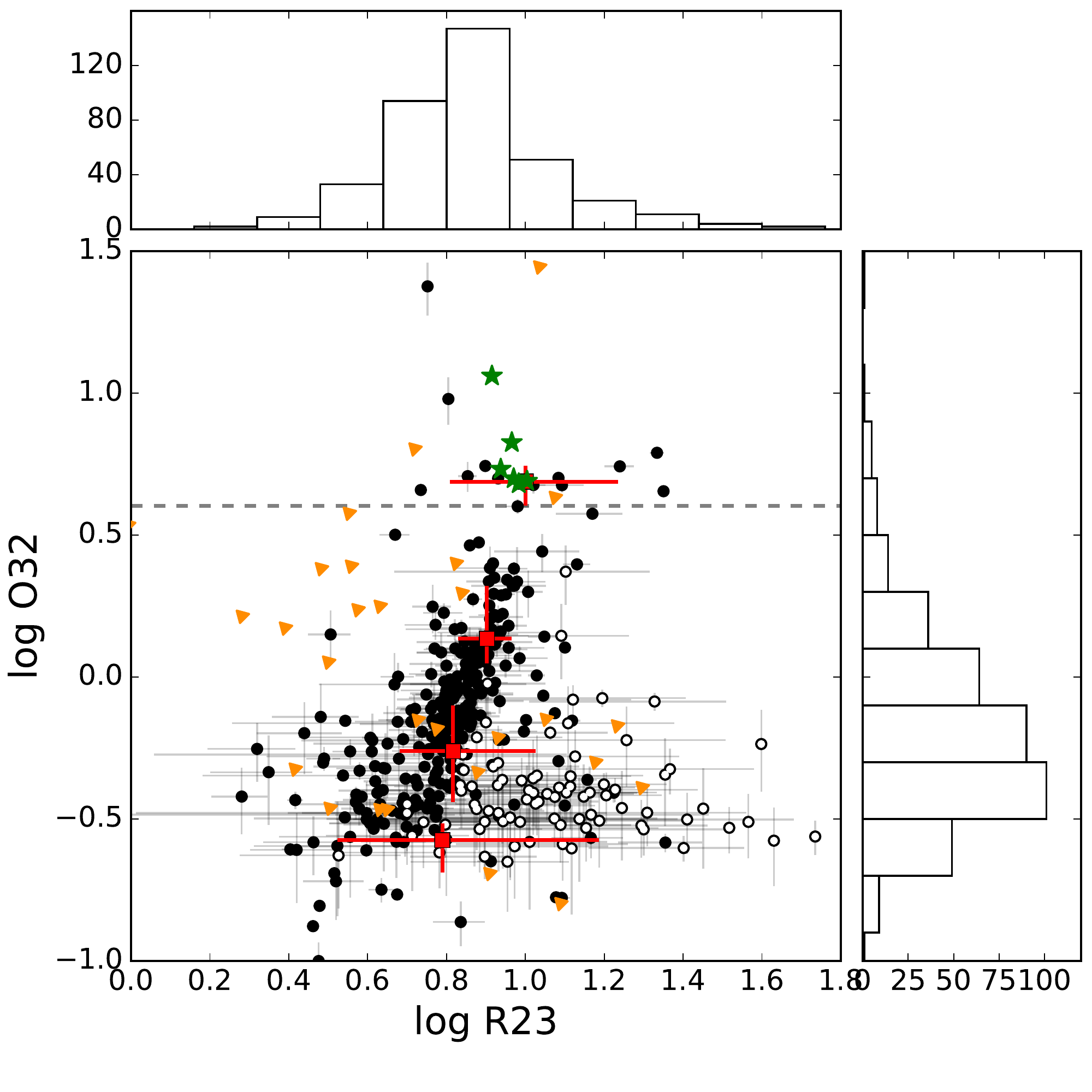}
   \caption{\ott\ versus \rtt , which is a proxy of the metallicity. The colours are used in the same way as in the previous plots. The orange triangles are now both lower limits of \ott\ and upper limits of \rtt . Galaxies with S/N(\hb) $< 3$ are shown by open symbols.} 
   \label{fig:r23}
   \end{figure}

\rtt\ is a diagnostic to estimate the gas-phase oxygen abundances (later referred to as metallicity, \Z ) of galaxies and is based on the ratio of \oii\ + \oiii\ over \hb\ \citep{1979A&A....78..200A, 1979MNRAS.189...95P, 1980MNRAS.193..219P, 1991ApJ...380..140M}, given by \citep{1991ApJ...380..140M}
\begin{equation}
\mathrm{\rtt} \space = \space \frac{ \oii{\ensuremath{\lambda}3727} +  \oiii{\ensuremath{\lambda}4959} +  \oiii{\ensuremath{\lambda}5007}}{\hb} \space. 
\end{equation} 
Since it only relies on the blue rest-frame spectrum, this diagnostic is often used when the red emission lines are out of the spectrum. However, the relation between \rtt\ and \Z\ is degenerate and therefore additional lines are still necessary to constrain the metallicity. In Fig. \ref{fig:r23} we plot the logarithm of the oxygen line ratio against the logarithm of \rtt . At high \ott\ ($\log$ \ott\ $\gtrsim$ -0.2), we visually determine a trend between \ott\ and \rtt , which is followed by the LyC leakers of \citet{Izotov16a,Izotov16b, 2018MNRAS.474.4514I}. However, below $\log$ \ott\ $\approx$ -0.2 the data points scatter in the $\log$ \rtt\ direction, as a result of uncertain \hb\ measurements as we will return to in the discussion. There we will also discuss how the stellar and gas-phase metallicity influences the oxygen ratio.

\section{Discussion}
\label{s_discussion}
In the previous section we gave an overview of the galaxy properties of our sample. Here we aim to determine what processes and properties are responsible for a high oxygen line ratio and under which circumstances it is likely to find extreme \ott\ ratios.  
\subsection{The occurrence of high \ott}
Although in previous studies 'extreme' oxygen emitters are defined as having \ott\ > 4, we use \ott\ > 1 as the threshold for 'high' \ott\ emitters to study their occurrence, which will be justified later in this section. This results in a significant sample of 104 high \ott\ emitters, in contrast to applying the \ott\ > 4 threshold, which would only leave 15 galaxies in the extreme regime. In order to study how our selection criterion influences our results and to study the redshift dependence of our results, we constructed a comparison sample from SDSS data. 
\subsubsection{Creating a comparison sample from SDSS data}
\label{section:sdsscomp}
We created a comparison sample of star-forming galaxies in SDSS DR7, from which the derivation of the measurements is detailed in \citet{2004MNRAS.351.1151B} and \citet{2004ApJ...613..898T}. For each galaxy in the MUSE sample we selected a local analogue in SDSS that lies at the same position in the redshift-corrected \SFR\ - \Mstar\ plane, thus both samples contain the same number of galaxies. However, because the spectra and consequently the emission-line flux measurement of SDSS galaxies are, unlike the MUSE galaxies, biased by aperture effects due to a finite fiber size, we first corrected for this. Assuming that the slope of the SFMS from Eq. \ref{SFMS} is unaffected, we refitted the relation to the SDSS data, resulting in a shift of +0.2 in the $\log \SFR - 2.93 \log(1+z)$ direction. Each SDSS galaxy in the same redshift-corrected \SFR\ - \Mstar\ position is a potential analogue of a MUSE galaxy. We therefore selected all SDSS galaxies within the $1- \sigma$ error bars of the position of the MUSE galaxy. We used the median value and the 16$\%$ and 84$\%$ percentiles of the \oiii\ and \oii\ emission-line fluxes of all selected SDSS galaxies galaxy as the fluxes and errors of the analogue.  

\begin{figure*}
 \centering
    \includegraphics[width=\hsize]{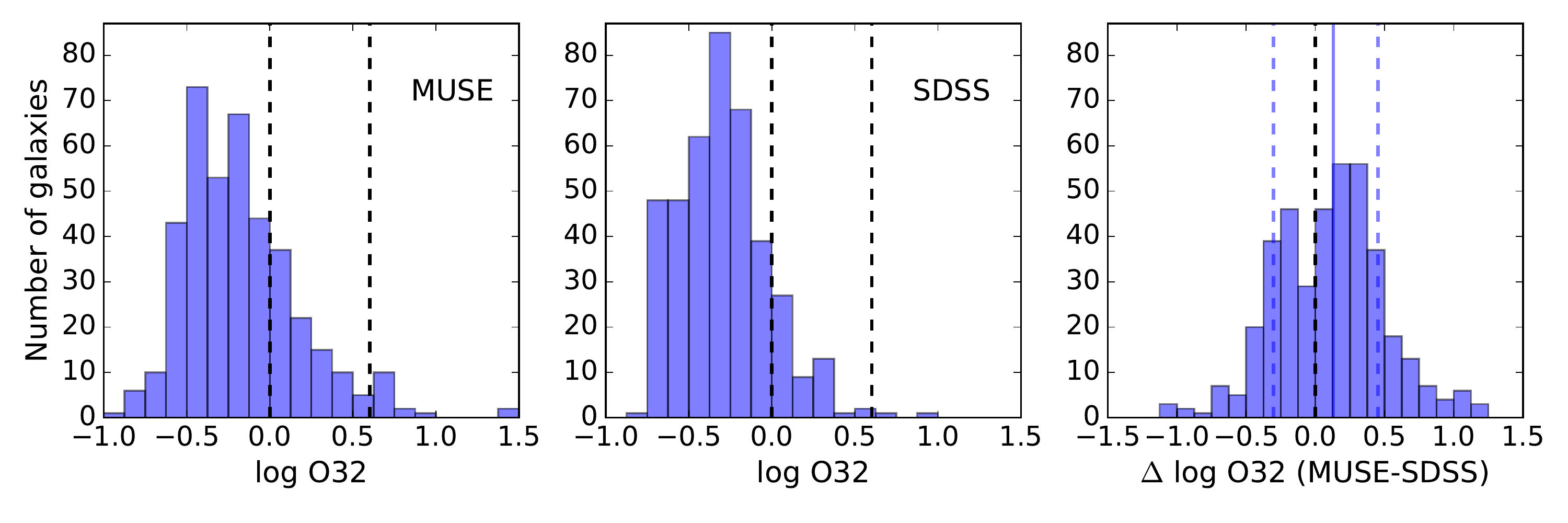}
      \caption{Number of galaxies per $\log$ \ott\ bin for our sample of 406 galaxies (left panel). The dashed lines correspond to \ott\ = 1 (left) and \ott\ = 4 (right). In the middle panel we show a similar plot for our SDSS comparison sample 1 (see the text for details). The distribution of the difference in \ott\ of each galaxy in our sample with its counterpart in the SDSS comparison sample is shown in the right panel. The median and the 1-$\sigma$ spread of the distribution are shown by the solid and dashed blue lines respectively.}
      \label{fig:oiii_oii_hist_sdss}
\end{figure*}

Figure \ref{fig:oiii_oii_hist_sdss} shows the distribution of \ott\ of our sample (left) and the SDSS comparison sample (middle). We compared the \ott\ of each galaxy in our sample with its counterpart in the SDSS sample. The result of this is shown in the right panel of Fig. \ref{fig:oiii_oii_hist_sdss}. At $\Delta \log$ \ott\ = 0 (black dashed line), the oxygen ratio of the galaxy in our sample is the same as its SDSS analogue. The median value of the $\Delta \log$ \ott\ is at 0.13 dex (blue solid line), which means that the \ott\ ratio of the MUSE galaxies on average exceeds the \ott\ of galaxies in the comparison sample. The dashed blue lines, however, indicate that this difference is within the 1-$\sigma$ error bars.

\subsubsection{Incidence rate of high \ott\ as a function of mass}
\label{incidenceratemass}
\begin{figure}
   \centering
   \includegraphics[width=\hsize]{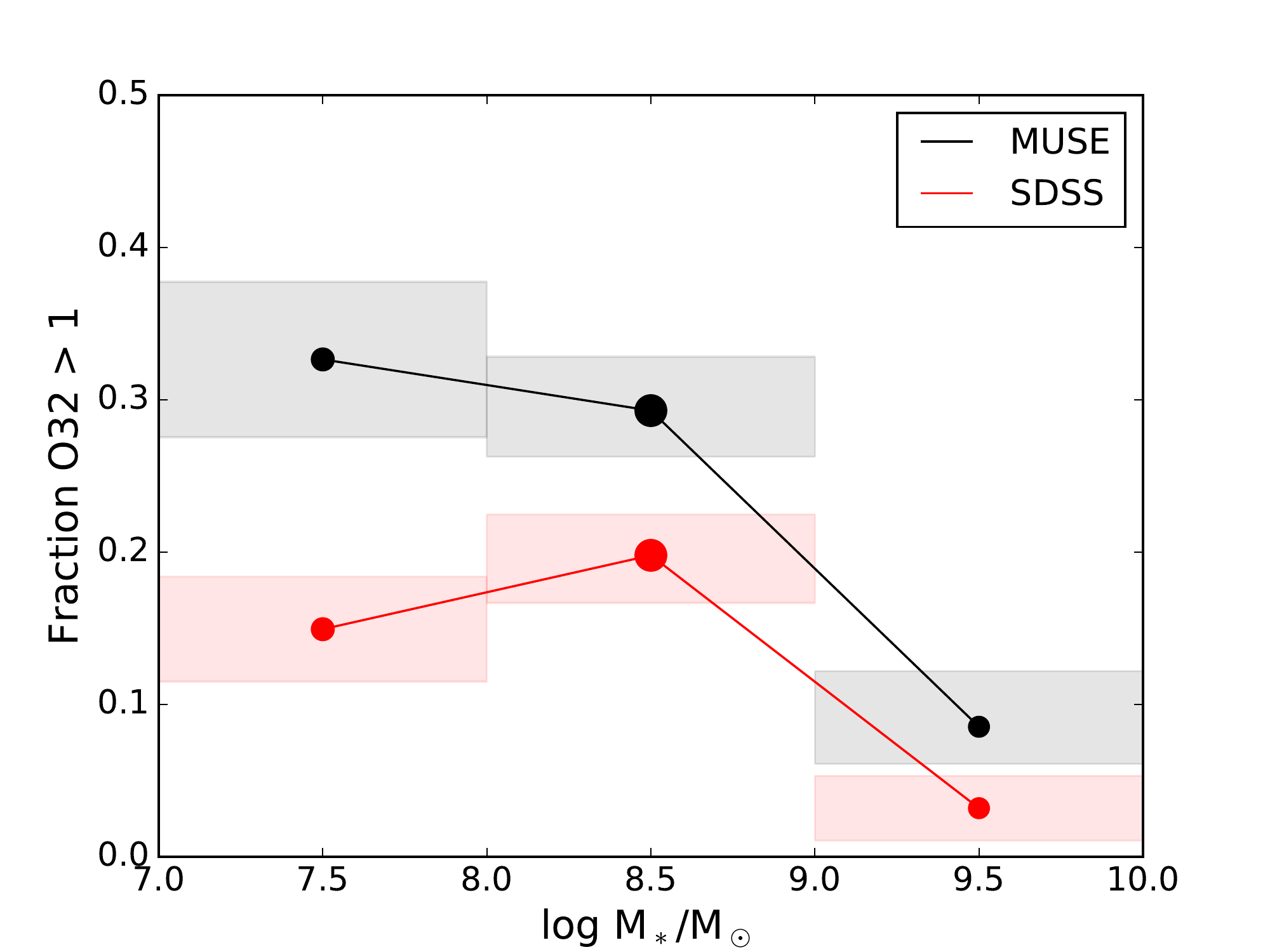}
   \caption{Fraction of galaxies with \ott\ > 1 in bins of 1 dex with central $\log$ \Mstar\/\Msun = 7.5, 8.5, and 9.5 of our MUSE sample (black/grey) and the SDSS comparison sample (red) (see the text for details). The points denote the centres of the stellar mass bins and their size reflects the number of galaxies in the bin. The shaded areas cover the 1-$\sigma$ errors as calculated by bootstrapping in the y-direction and the bin size in the x-direction.} 
   \label{fig:f_logM_notZcorr}
   \end{figure}

We divided the sample into three subsets based on stellar mass, with $\log$ \Mstar /\Msun\ in the ranges  [7.0, 8.0],  [8.0, 9.0], and  [9.0, 10.0]. We then derived the fraction of galaxies with \ott\ > 1 in each mass bin, shown as the black line in Fig. \ref{fig:f_logM_notZcorr}, where the points indicate the centres of the mass bins. One-$\sigma$ errors are derived by bootstrapping the sample 10,000 times (grey area). We then applied the same approach to the SDSS comparison sample (see the red points, line, and shaded area).

The fraction of galaxies in the MUSE sample with \ott\ > 1 decreases with increasing \Mstar\ (black line); we find $\sim$30 $\%$ for galaxies with stellar masses between $10^{7}$ and $10^{9}$ \Msun, but $\sim$10 $\%$ in the highest mass bin. This trend is comparable to that of the median bins between \ott\ and \Mstar\ , which we described in Sect. \ref{subsec:o32M} and Fig. \ref{fig:m_o32}. The SDSS comparison sample follows a comparable trend, but the fractions are offset by 0.05-0.2 towards lower fractions. Below $\log$ \Mstar/\Msun\ $ \approx$ 8 the SDSS sample is, however, incomplete (see Appendix \ref{app:sdss}), so we caution that the SDSS results in the lowest mass bin are likely to be biased as a result. 

\subsubsection{Adopting a metallicity-dependent threshold for the \ott\ ratio}
\label{sec:metalthreshold}
\begin{figure}
   \centering
   \includegraphics[width=\hsize]{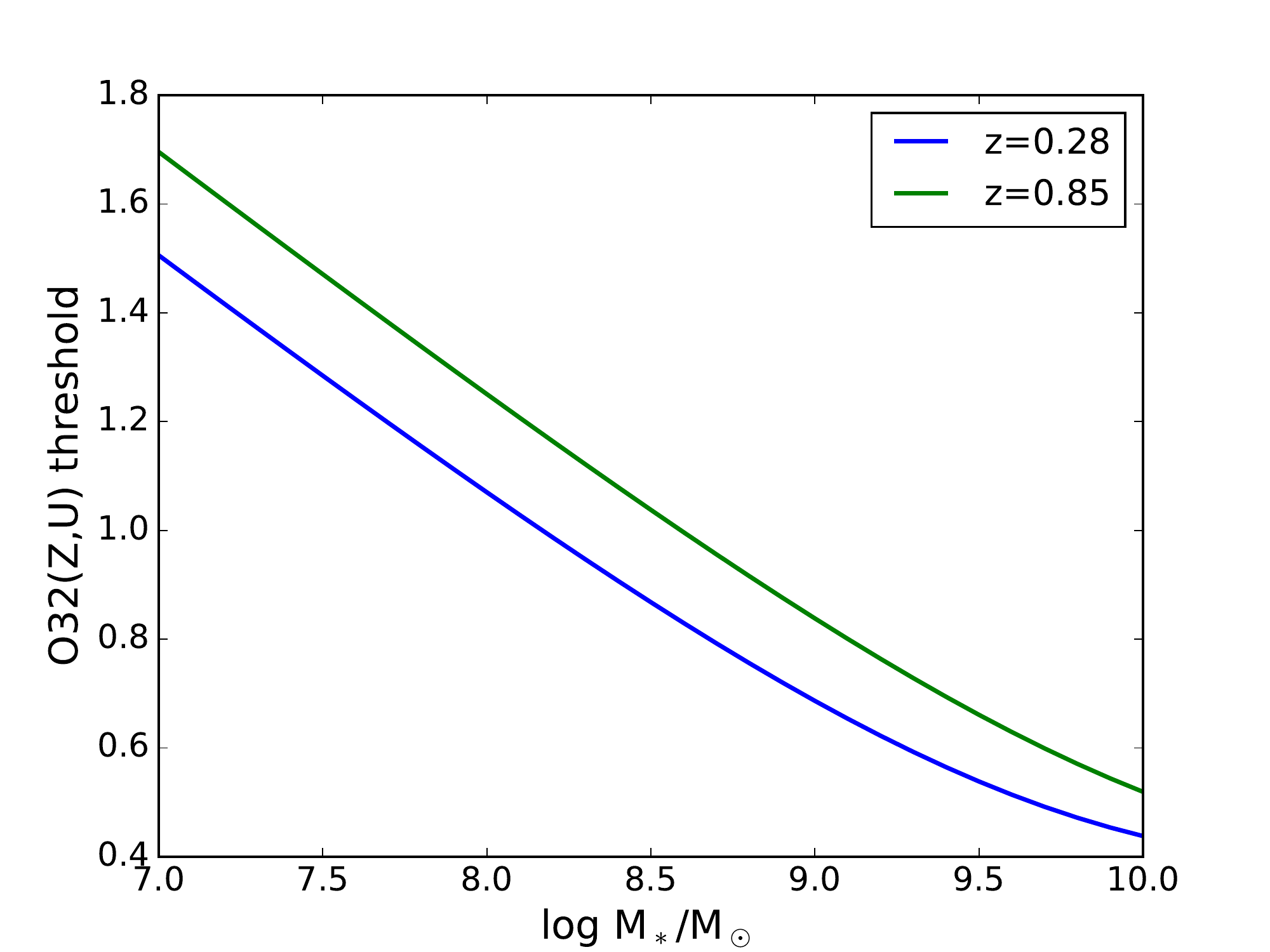}
   \caption{Metallicity-dependent \ott\ threshold as a function of stellar mass, for the minimum and maximum redshift of our sample, derived using the redshift-dependent \Mstar\ - \Z\ relation of \citet{2014ApJ...791..130Z} and the \ott\ ratio from photo-ionisation models from \citet{2016MNRAS.462.1757G} (see the text for details).
   }
   \label{fig:threshold}
   \end{figure}

\begin{figure}
   \centering
   \includegraphics[width=\hsize]{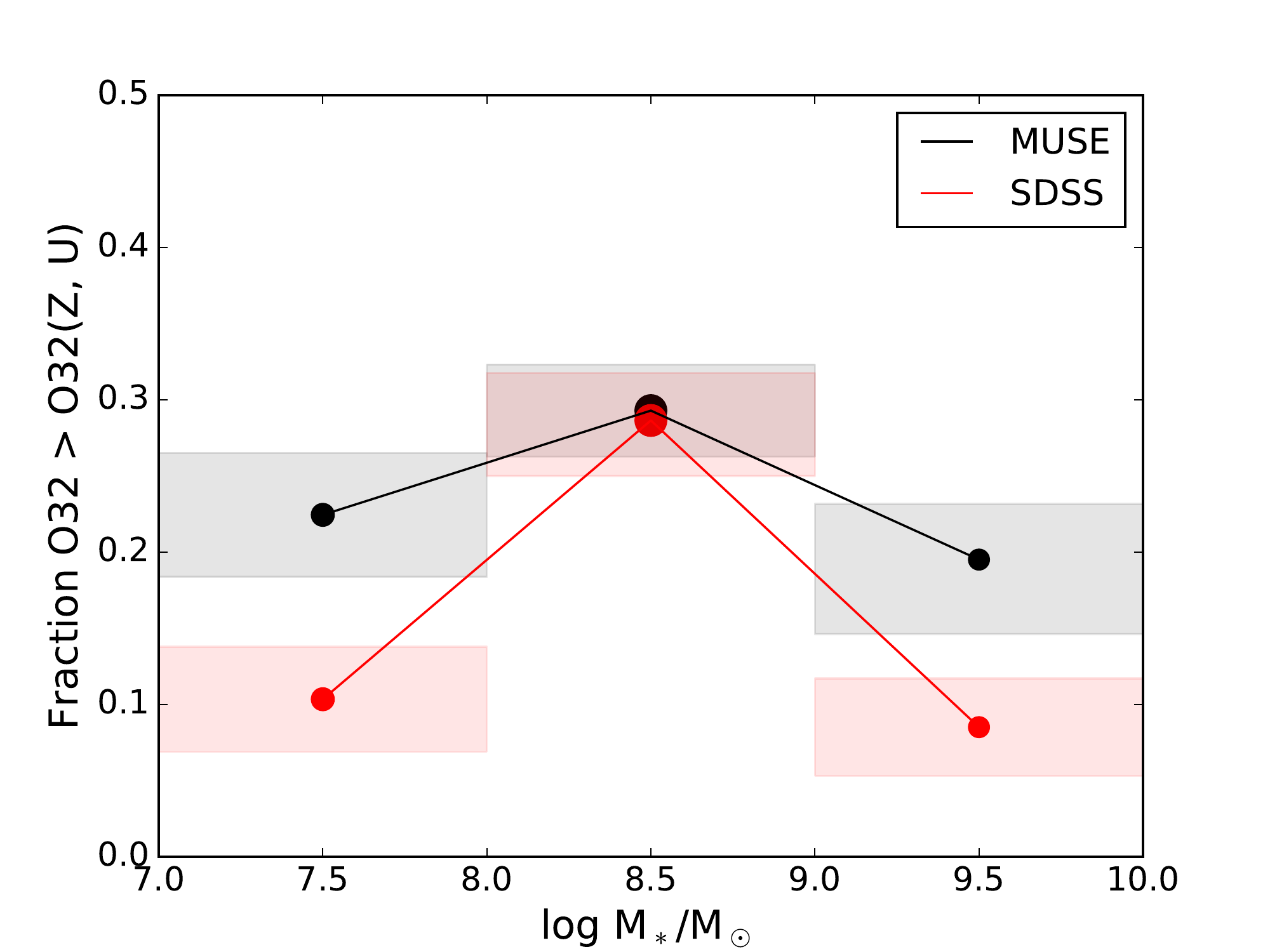}
   \caption{Fraction of galaxies with \ott\ greater than the metallicity-dependent threshold for our MUSE sample (black/grey) and the SDSS comparison sample (red) (see the text for details).} 
   \label{fig:f_logM_Zcorr}
   \end{figure}

High oxygen ratios can also be driven by low metallicity systems (see for example the nebular models of \citealt{2016MNRAS.462.1757G}) due to the harder ionising spectrum of low metallicity stars and less efficient cooling. Here we perform the same analysis as in the previous section, but instead of the fixed threshold at \ott\ = 1, we adopt a metallicity dependent threshold on \ott\ that we derive as follows. For each galaxy we derive the metallicity \Z\ using the redshift dependent \Mstar\ - \Z\ relation of \citet{2014ApJ...791..130Z}. We then set the threshold for this galaxy equal to the \ott\ ratio from photo-ionisation models from \citet{2016MNRAS.462.1757G} with this metallicity and the ionisation parameter set to $\log U = -3$. We show the relation between the metallicity dependent \ott\ threshold and stellar mass in Fig. \ref{fig:threshold},  for the minimum and maximum redshifts of our sample ($z = 0.28$ and $z = 0.85$). Our assumption that galaxies are highly ionised above $\log U = -3$ results in an \ott\ threshold for \ott\ between 0.5 and 1.7. This is the reason for setting the fixed threshold to \ott\ > 1  in the previous section. The incidence rate of galaxies with \ott\ above this metallicity-dependent threshold in each mass bin is shown in Fig. \ref{fig:f_logM_Zcorr} for the MUSE sample (black/grey) and the comparison sample (red).

We see that with the \Z -dependent threshold there is no longer a strong trend between \Mstar\ and the fraction of high \ott\ emitters. For the bin with most galaxies ($8 < \log$ \Mstar/\Msun\ $< 9$, see Table \ref{table:fractions}),  the SDSS and MUSE results are in agreement. This indicates that the relation between \Mstar\ and the incidence rate of high \ott\ emitters is most likely the result of the relation between metallicity and oxygen ratio.

\subsubsection{Evolution of the fraction of galaxies with high \ott}
\begin{figure}
   \centering
   \includegraphics[width=\hsize]{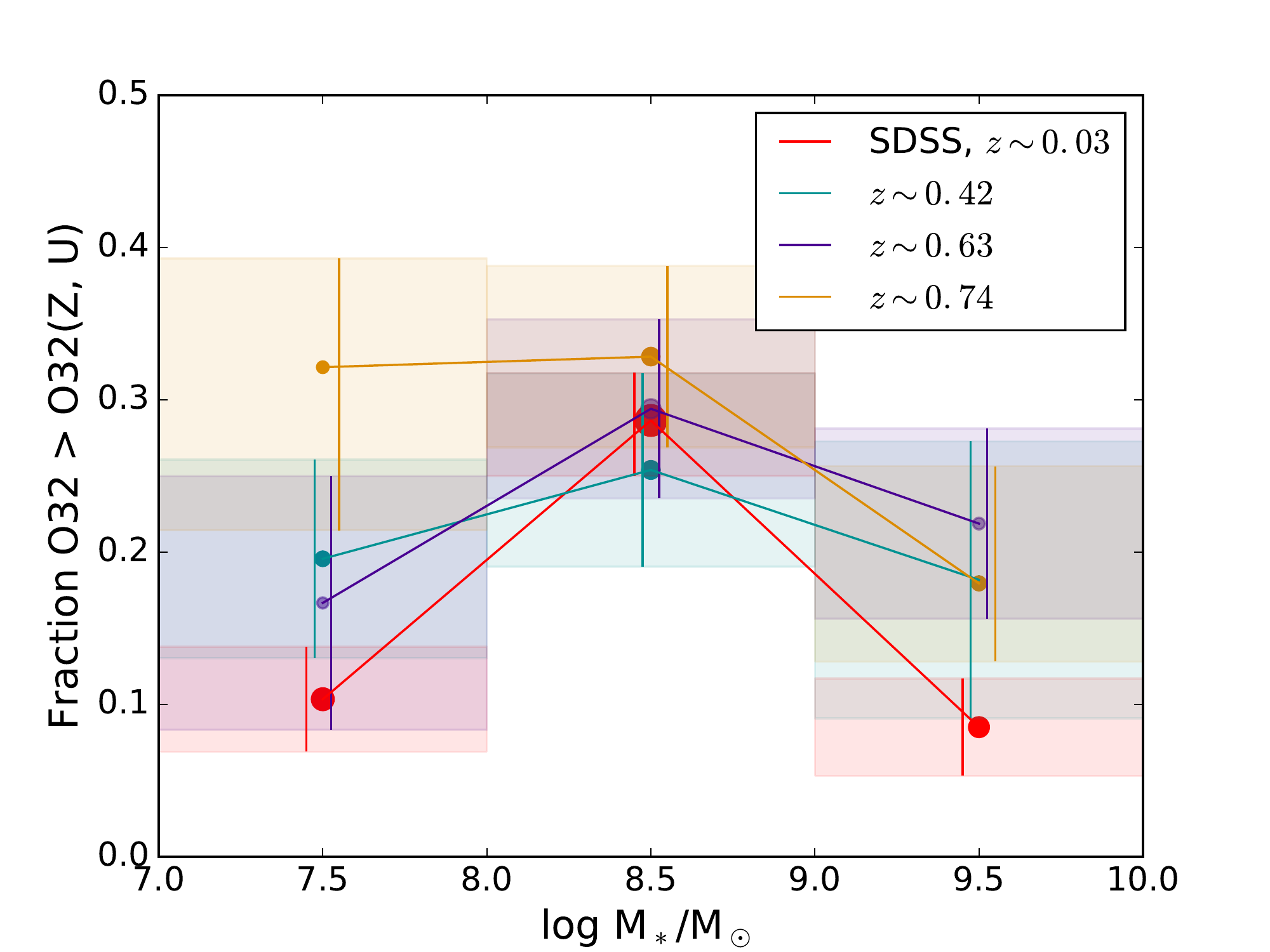}
   \caption{Incidence rate of galaxies with \ott\ above the metallicity-dependent threshold for three equally sized redshift-selected subsets, with median redshifts of z = 0.42 (green), z = 0.63 (purple), and z = 0.74 (orange) and the entire SDSS comparison sample (red) with median redshift z = 0.03. The vertical lines reflect the 1-$\sigma$ errors in each bin.} 
   \label{fig:f_logM_Zcorr_zdep}
   \end{figure}

We split the sample into three redshift bins of equal number, with median redshifts of $z = 0.42$, $z = 0.63,$ and $z = 0.74$. We then calculated the fractions of galaxies above the metallicity-dependent threshold and show the result in Fig. \ref{fig:f_logM_Zcorr_zdep}. For comparison we also show the incidence rates of the entire SDSS comparison sample, which has a median redshift of $z = 0.03$. 

\begin{figure}
   \centering
   \includegraphics[width=\hsize]{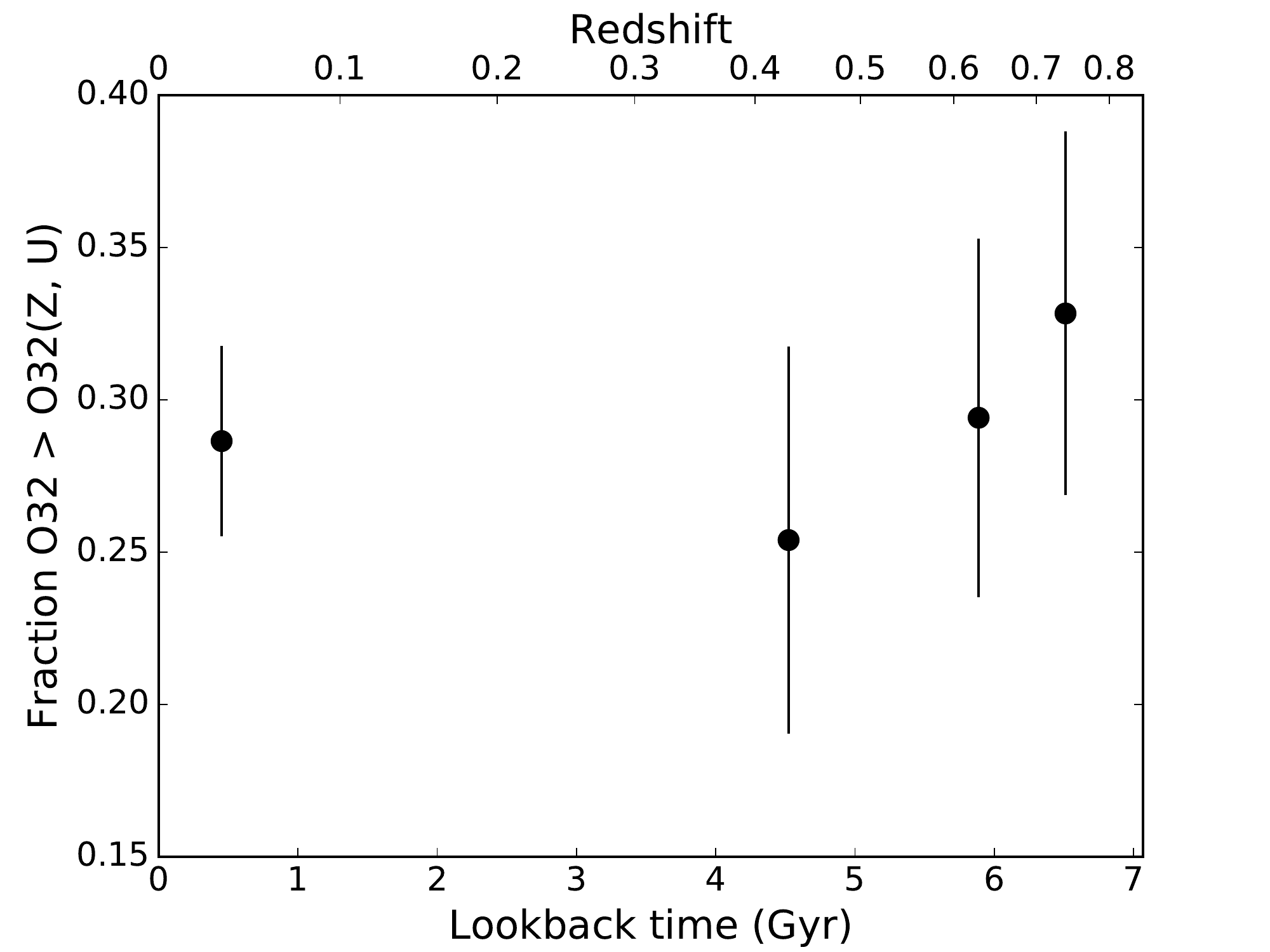}
   \caption{Incidence rate of galaxies with high oxygen emission for galaxies with stellar masses between $\log$ \Mstar /\Msun\  = 8 and $\log$ \Mstar /\Msun\  = 9, versus look-back time and redshift, calculated from the median redshift of the subsets.} 
   \label{fig:fraction_time}
   \end{figure}

In the lowest mass bin there seems to be a weak trend between the incidence rate and the redshift, for example the fraction in the highest $z$ bin is significantly higher than the fraction of the SDSS comparison sample. However, the lowest redshift subsample in this mass bin is larger (44) than the intermediate and high-redshift bin (26 and 28 respectively; see also Table \ref{table:fractions}), which may indicate that we only include the most extreme star-forming systems in the high-redshift sample and this can explain the results that we observe in the lowest mass bin. In the two highest mass bins, there is no significant difference between the fraction of \ott\ > 1 at different redshifts. In Fig. \ref{fig:fraction_time} the results for the galaxies with stellar masses between $\log$ \Mstar /\Msun\  = 8 and $\log$ \Mstar /\Msun\  = 9 are presented against the look-back time and redshift, which we calculated using the median redshifts of the redshift bins. 
   
The incidence rate of high oxygen emitters when controlled for metallicity is thus independent of redshift for our sample. Since we selected the galaxies based on their distance to the redshift-dependent main SFMS (Eq. \ref{SFMS}), we selected the same fraction of star-forming galaxies at each redshift, rather than selecting the same kind (same \SFR\ and \Mstar\ and thus \sSFR ) of galaxies over cosmic time. The incidence rates that we here derived may therefore be extrapolated to the entire population of star-forming galaxies, and suggest that the fraction of high \ott\ emitters are constant over time between $z = 0.28$ and $z = 0.85$. 

In the current paradigm, the \ott\ ratio of high-redshift galaxies are believed to be more extreme than those of low-redshift galaxies. Results from the MOSFIRE Deep Evolution Field (MOSDEF) and the Keck Baryonic Structure Survey (KBSS)-MOSFIRE surveys of galaxies at $z \sim 2.3$ show that their \ott\ ratios are offset towards significantly larger values in comparison to those of local galaxies \citep{2014ApJ...795..165S, 2016ApJ...816...23S, 2017ApJ...836..164S}. This may be interpreted by a harder stellar radiation field at fixed mass at higher redshift \citep{2014ApJ...795..165S, 2017ApJ...836..164S}, or as the result of lower metallicities of high-redshift galaxies at fixed mass \citep{2016ApJ...816...23S}. These results are supported by cosmological simulations of massive galaxies \citep{2017MNRAS.472.2468H} that show that the ionisation parameter and the \oiii/\hb\ ratio increases with redshift at fixed \Mstar\ at $0 < z < 4$. Our results at $0.28 < z < 0.85$ , however, support another scenario where the \ott\ ratio is constant over cosmic time. This difference may be the result of the different \Mstar\ regime that is probed in the high-redshift surveys (9 $\lesssim$ $\log$ \Mstar/\Msun\ $\lesssim$ 11) and in the cosmological simulations (9.5 $\lesssim$ $\log$ \Mstar/\Msun\ $\lesssim$ 11.5) with respect to the stellar mass of the galaxies in this work (7 $\lesssim$ $\log$ \Mstar/\Msun\ $\lesssim$ 10).    \begin{table*}
\caption{Number of galaxies in mass and redshift bins.}             
\label{table:fractions}      
\centering                          
\begin{tabular}{c | c c c c}        
\hline\hline                
 & 7< $\log$ \Mstar /\Msun\ <8 & 8 < $\log$ \Mstar /\Msun\ < 9 & 9 < $\log$ \Mstar /\Msun\ <10 & total \\
  \hline                        
all & 98 & 195 & 81 & 406 \\
low z & 44 & 64 & 11 & 135 \\
intermediate z & 26 & 67 & 29 & 135 \\
high z  & 28 & 64 & 41 & 136 \\
\hline                                   
\end{tabular}
\end{table*}

\subsubsection{Completeness and robustness of the results}
Our data sample consists of a combination of several surveys with depths between one and 30 hours. The catalogues for most fields are a mixture of emission lines and continuum-detected galaxies, resulting in samples of different completeness limits in \Mstar\ and emission-line flux. In Appendix \ref{app:sdss} we study how such a possible incompleteness of our data sample alters our results by simulating different completenesses in flux and \Mstar\ of SDSS data, and find that the effect is negligible. 

We studied the impact of our \SFR\ calibration method on the results and compared them with \hb -derived \SFRs\ that are de-reddened by the \hb/\hg\ fraction. We also re-analysed the data by adopting different stellar libraries for \Mstar\ calculations and derived similar results. However, for one of the surveys that is used for this study, the MUSE QuBES, we used the MUSE spectrum for the SED fitting instead of deep photometry as we used for the data of the other surveys. We are aware that this induces uncertainty on the mass estimates and therefore re-analysed the results in Sect. \ref{sec:metalthreshold} without the data of the MUSE QuBES survey and acquired comparable results.

\subsection{Can nebular models with no escape of ionising photons predict the observed \ott ?}
\begin{figure*}
   \centering
   \includegraphics[width=\hsize]{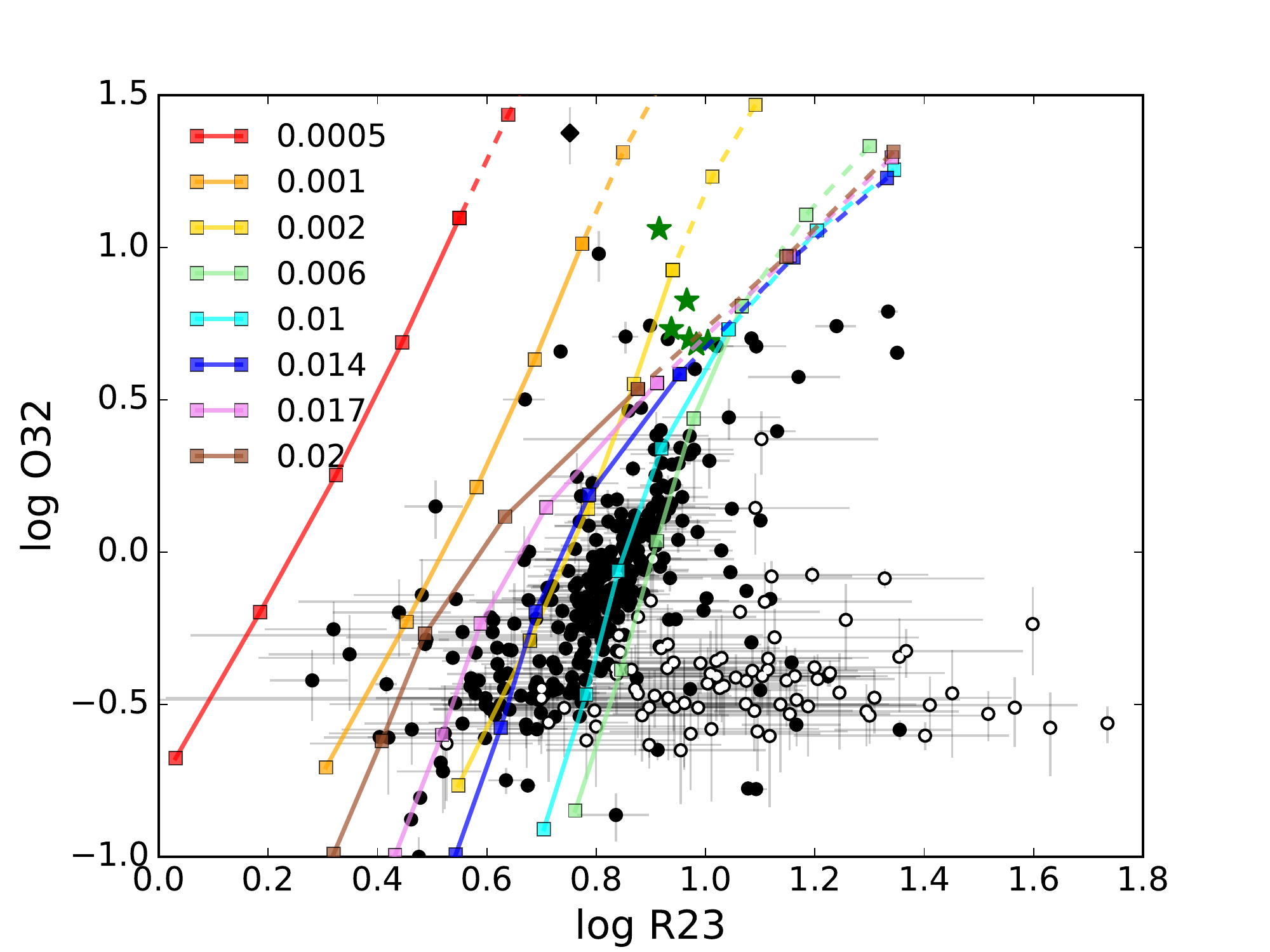}
   \caption{Logarithm of \ott\ versus the logarithm of \rtt . The black points are galaxies from our sample, open symbols reflect galaxies with S/N (\hb ) < 3), while the green stars are the confirmed LyC leakers from \citet{Izotov16a,Izotov16b, 2018MNRAS.474.4514I}. The squares that are connected by solid ($\log U < -2$) and dashed coloured lines show the position in this diagram of the nebular model of \citet{2016MNRAS.462.1757G} for constant gas-phase metallicity (see legend) and with increasing $\log U$ towards the upper right, with -4 < $\log U$ < -1 and step size 0.5.} 
   \label{fig:r23_discussion}
   \end{figure*}

In Sect. \ref{results:r23} we discussed the behaviour of our galaxies in the $\log$ \ott\ versus \rtt\ diagram (see also Fig. \ref{fig:r23}). Here we compare these results with nebular models to study if they are consistent with each other and whether we can derive the nebular metallicity of our galaxies with the \rtt\ method (see Fig. \ref{fig:r23_discussion}). We added the grids of line ratios from nebular models that are calculated by \citet{2016MNRAS.462.1757G} by the coloured squares, where each colour represents a model of a fixed metallicity \Z \ (see the figure legend; metallicity here refers to the combination of nebular and stellar oxygen abundances, since these are kept constant for these models). We connected the grids of models with constant metallicity by dashed lines, where the ionisation parameter $U$ increases towards the upper right from $\log U = -4$ to $\log U = -1$. For the calculation of these models the ionising photon escape fraction was assumed to be zero. The lower limits on \ott,\ which are also upper limits on \rtt,\ are not shown here since in this plot because it is difficult to compare these galaxies with models. 

Deriving the nebular metallicity of our galaxies by comparing it to the model results is not straightforward, due to the degeneracy of the \rtt--metallicity relation. However, from Fig. \ref{fig:r23_discussion} it is clear that the \rtt\ ratio indicates that the metallicity of the majority of the galaxies in our sample is sub-solar and around 0.006 ($\approx 1/3$ \Zsun). Hence the \rtt\ of many of the galaxies exceeds the maximum \rtt\ predicted by the models, which can partly be explained by an uncertain \hb\ flux measurement (galaxies with an \hb\ signal to noise lower than three are shown by open squares). 

We have differentiated between models with ionisation parameters between  $\log U$ = -4 and $\log U$ = -2 (solid line) and those of higher values, since observations show that the bulk of star-forming galaxies have $\log U$ < -2 (e.g. \citealt{2014ApJ...787..120S}). The ionisation parameter of galaxies with extreme \ott\ ratios might however exceed those of normal galaxies, as pointed out by \citet{Stasinska15}.  However, if the \ott\ ratio of our galaxies exceeds the predicted ratio of models with $\log U > -2$, the escape of LyC photons is also a likely scenario. For this reason the galaxy with the highest \ott\ ratio is a promising LyC escape candidate. Although the \ott\ ratios of the other extreme emitters in our sample (with \ott\ > 4) in this diagram are similar to those of the confirmed LyC leakers (green stars), the logaritm of \rtt\ of our galaxies scatters within 0.4 of the value of the LyC leakers. Most of them however imply either low stellar and nebular metallicities, high ionisation parameters, the escape of ionising photons, or a combination of these factors. However, comparing the data that lie inside the model grid to these nebular models with no escape of ionising photons is thus not sufficient to determine if LyC escape is responsible for the extreme oxygen emission. 
   
\subsection{Extreme \ott\ emitters}
In the previous sections we discussed how the properties of galaxies with extreme oxygen ratios are related to the entire sample. Here we study the robustness of the \ott\ measurement and investigate the electron temperatures of the \oiii\ regime in extreme galaxies. 

\begin{figure}
   \centering
   \includegraphics[width=\hsize]{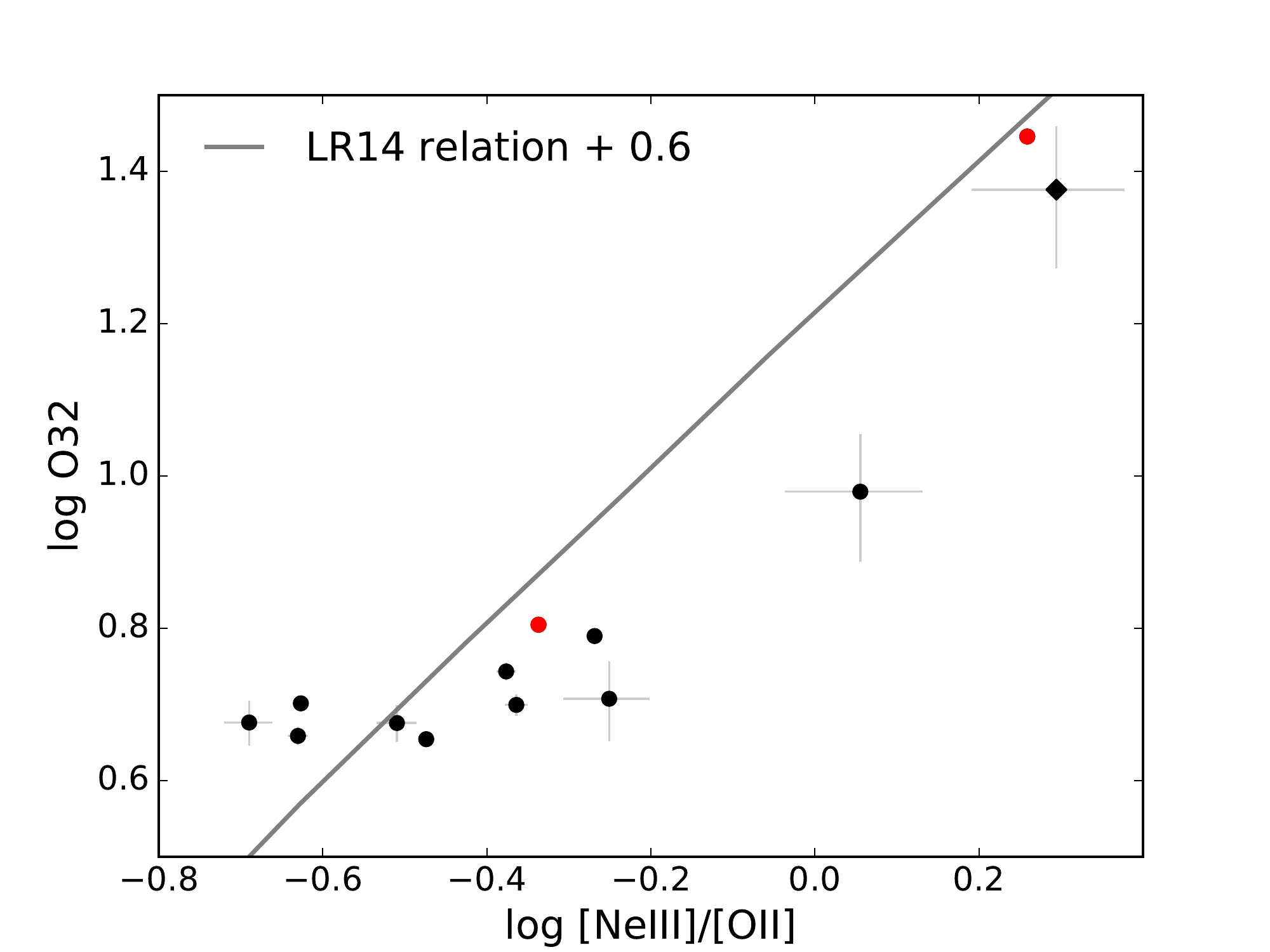}
   \caption{Log \ott\ versus the $\log$ \neiiil/\oii\ ratio for our extreme emitters (\ott\ >4) with  a \neiiil\ detection of at least 3 $\sigma$. The galaxies of which we do not have a significant \oii\ detection are presented by red circles, which are the 3-$\sigma$ upper limits on both ratios. The rest of the extreme emitters sample is shown by black circles and the most extreme \ott\ emitter by the black diamond. The grey line corresponds to the \citet{2014ApJ...780..100L} relation for \Z\ = 0.2 \Zsun\ between $\log$ \ott\ and $\log$ \neiiil/\oii\ , which is offset by +0.6 in the y-direction.} 
   \label{fig:ne3}
 \end{figure}

To confirm that the \ott\ ratios are well measured, we compare them to the \neiiil/\oii\ ratio. This ratio is an alternative diagnostic of the ionisation parameter because of its tight relation to the \ott\ ratio \citep{2014ApJ...780..100L}. Even though \neiiil\ is more than an order of magnitude fainter than \oiiil , the \neiii/\oii\ ratio is less affected by reddening than the \ott\ ratio. 
The relationship between  \oiiil/\oiil\ and \neiiil/\oiil\  for the extreme emitters with a significant \neiiil\ detection ($\sigma > 3$) is shown in Fig.~\ref{fig:ne3}. The solid grey line shows the predictions of the Starburst99/Mappings III photo-ionisation models of \citet{2014ApJ...780..100L} with \Z\ = 0.2 \Zsun . We offset the models by a factor of +0.6 in the $\log$ \oiiil/\oiil\ direction to take into account the discrepancy between these models and the observations of  \citet{2006A&A...448..955I} and \citet{Jaskot13}, as discussed by \citet{2014ApJ...780..100L}. 
 We use these results to conclude that the \neiiil/\oii\ ratios are consistent with extreme \ott\ ratios. There are, however, offsets between our observations and the corrected \citet{2014ApJ...780..100L} relation of up to 0.2 dex, which are most likely caused by the lower significance of the \neiiil\ line and/or by an offset in the dust attenuation estimate. We note however, that if we use these offsets to correct the \oiii\ line fluxes,  the \ott\ ratios will still be in the regime of the extreme oxygen emitters.

\begin{figure}
   \centering
   \includegraphics[width=\hsize]{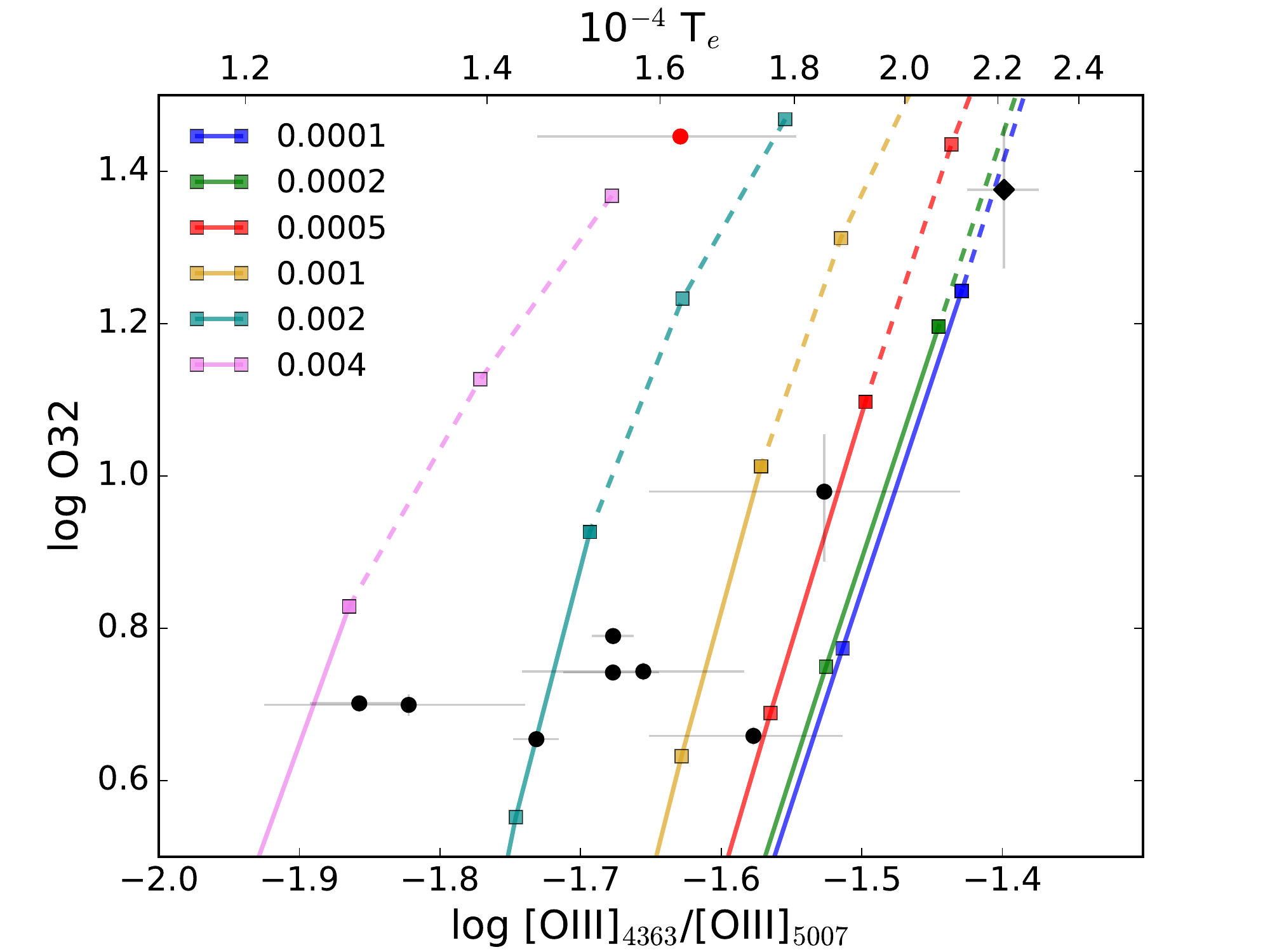}
   \caption{Log \ott\ versus the \oiiia/\oiiil\ ratio (lower x-axis) and the electron temperature in the \oiii\ regime (lower x-axis)  calculated using the ${\tt nebular.ionic}$ routine in Pyneb \citep{2015A&A...573A..42L}, assuming $n_e$ = 100 cm$^3$. The coloured lines indicate the predictions from the \citet{2016MNRAS.462.1757G} models of different metallicity as indicated by the colours, with increasing $\log U$ towards the upper right, and $\log U < -2$ (solid line) and $\log U < -2$ (dashed line), and step size of $\log U$ = 0.5 between the squares. The colours of the circles are used as in Fig. \ref{fig:ne3}. }
   \label{fig:te}
 \end{figure} 
 For most of the extreme emitters, we have high S/N measurements (S/N > 3) of the auroral \oiiia\ emission line. As such we can use the \oiiia/\oiiil\ ratio to diagnose the electron temperatures (\Te) in the \oiii\ regime of the \hii\ region (see \citealt{1989agna.book.....O}), using the ${\tt nebular.ionic}$ routine in Pyneb \citep{2015A&A...573A..42L}, assuming an electron density of $n_e = 100$ cm$^{-3}$.
The \oiiia/\oiiil\ ratios and the corresponding values of \Te\ are shown in Fig. \ref{fig:te}. A complete discussion on this is outside the scope of the paper. However, there is presumably a positive correlation between \ott\ and \Te\ since the electron temperatures of the galaxies with the most extreme \ott\ ratios exceed those of the other extreme galaxies. 
Furthermore, comparing the \oiiia/\oiiil\ and \ott\ ratios to the predictions from the \citet{2016MNRAS.462.1757G} nebular models, the metallicity of the galaxies can be constrained, varying from \Z$\approx0.0001$ (\Z/\Zsun$\approx$0.005) for the galaxy with the highest \ott\ ratio (diamond) to \Z$\approx0.002-0.004$ (\Z/\Zsun$\approx$0.15) for the galaxies with \ott $\approx 4$. Again, for the calculation of these nebular models, the fraction of escaping ionising photons is assumed to equal zero. Although such a scenario would boost the \ott\ ratio, there is no obvious reason to expect an effect on the \oiiia/\oiiil\ ratio, and thus the real metallicity of a leaking system would be even lower than what is predicted by the models. Other scenarios, such as a top-heavy initial mass function, might increase the hardness of the ionising spectrum and could, however, also affect the position of a galaxy in this plot. 
 
\subsubsection*{UDF object 6865}
\begin{figure} 
   \centering
   \includegraphics[width=\hsize]{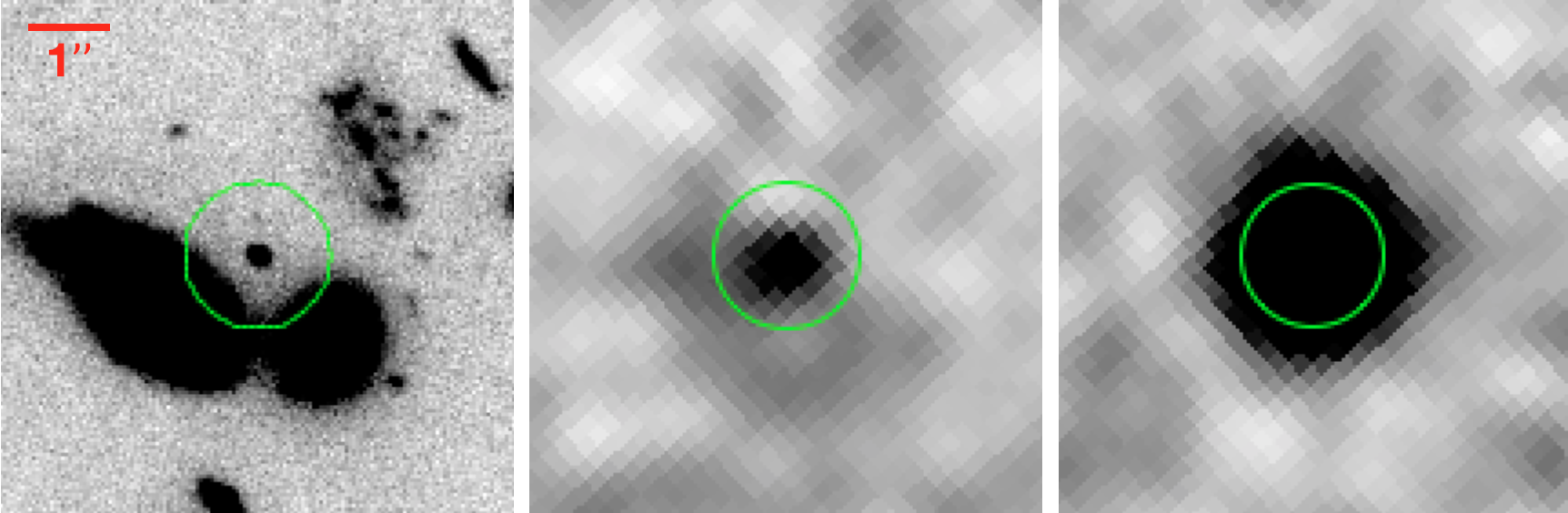}
   \caption{HST F77W image (left), a MUSE narrow band image of the \oiil\ line (middle), and a MUSE narrow band image of the \oiiil\ line (right) of the most extreme oxygen emitter with \ott\ = 23.} 
   \label{fig:images}
   \end{figure}

In Fig. \ref{fig:images} we show an HST image of the F775W filter and MUSE images of the \oii\ and \oiii\ lines of the most extreme oxygen emitter in our sample, which has \ott\ = 23. This galaxy is observed in the UDF-mosaic at $z = 0.83$ (identified in the MUSE catalogue of \citealt{2017A&A...608A...2I} as id = 6865). The spectrum of this object is shown in the upper panel of  Fig. \ref{fig:example_spec}, from which we derived an extinction \tauv\ =  0.49. The field is very crowded as can be seen in the HST image. The MUSE narrow band images of the \oiil\ and \oiiil\ lines, however, confirm that the measured line fluxes originate from this source in the centre of the images. The $\log$ \rtt , $\log$ \neiii/\oii,\ and  $\log$ \oiiia/\oiiil\ of this object are shown by the diamond in Figs. \ref{fig:r23_discussion}, \ref{fig:ne3}, and \ref{fig:te}. The $\log$ \oiiia/\oiiil\ indicates that the metallicity of this galaxy is extremely low (\Z$\approx0.0001$), which deviates somewhat from what is predicted from the \rtt\ ratio (\Z$\approx0.0005-0.001$), and the ionisation parameter is high ($\log U \approx -1.5$). However, all models assume that there is no escape of ionising photons, which may affect the \ott\ and \rtt\ ratios. 

Due to the combination of the relatively low stellar mass, $\log$(\Mstar/\Msun)$=8.16^{+0.16}_{-0.06}$, and the redshift, we are not able to precisely derive the size of the object, because the apparent size is comparable to the PSF of the HST image and the source is not resolved in the MUSE data. This, however, indicates that the galaxy is compact, like the objects of \citet{Izotov16a,Izotov16b, 2018MNRAS.474.4514I}. Together with the comparison of its oxygen ratio to those of nebular models, this suggests that this galaxy may be an LyC emitter candidate. 

\section{Conclusions}
We constructed a sample of emission-line galaxies in the redshift range $0.28 < z < 0.85$ that are detected in data from four MUSE GTO surveys. The galaxies are selected based on their position in the \SFR\ - \Mstar\ plane in a way that we only included galaxies that are above the redshift-dependent SFMS from \citet{Boogaardetal}. In this regime we expect the sample to be independent of selection effects. Our final sample consists of 406 galaxies, of which 104 (26$\%$) have a high \ott\ ratio (\ott\ > 1) and 15 galaxies are extreme emitters with \ott > 4 (3.7$\%$). We studied the \ott\ ratio as a function of the position in the (redshift-corrected) \SFR\ versus \Mstar\ diagram, as a function of stellar mass \Mstar , \SFR , and metallicity indicator \rtt . We then studied the incidence rate of galaxies with high oxygen ratios, which is defined by either a fixed threshold, \ott\ > 1, or by a metallicity-dependent threshold as a function of \Mstar\ and redshift. The main conclusions of this study are: 
\begin{itemize}
\item Galaxies with a high oxygen ratio are more common at lower masses (\Mstar\ < 9) and above the SFMS. There is no clear correlation between distance from the SFMS and the \ott\ ratio for galaxies in our final sample that are above the SFMS (Fig. \ref{fig:main_sequence}). \\
\item We find no correlation between \ott\ ratio and \Mstar , although the median values in \ott\ bins seems to be anti-correlated (Fig. \ref{fig:m_o32}).\\
\item We observe the same trend between the median values of \ott\ and  \SFR , but again no significant correlation for individual galaxies. The \SFR\ of most of our extreme emitters is two to three orders of magnitude smaller than those of confirmed leakers (Fig. \ref{fig:sfr_o32}).\\
\item The fraction of galaxies with high \ott\ ratios is independent of stellar mass when we use a metallicity-dependent \ott\ threshold (Fig. \ref{fig:f_logM_Zcorr}).\\
\item We find no significant correlation between the fraction of high \ott\ emitters and redshift, suggesting that there is no redshift evolution of the number of high \ott\ in the redshift range $0.28 < z < 0.85$ (Figs. \ref{fig:f_logM_Zcorr_zdep} and \ref{fig:fraction_time}).  \\

\item Comparing \ott\ and \rtt\ of our galaxies with those of nebular models with no escape of ionising photons, we find that some of the high oxygen emitters can be reproduced by models with a high ionisation parameter ($\log U \approx -2$), a very low stellar and nebular metallicity (smaller than $\sim 1/3$ \Zsun), or a combination of both. However, our extreme emitters are in the same regime as the confirmed leakers from \citet{Izotov16a, Izotov16b, 2018MNRAS.474.4514I} and we therefore cannot exclude the escape of ionising photons from these galaxies. The \ott\ ratio of our most extreme oxygen emitter can only be explained by models with very high ionisation parameter ($\log U > -2$), from which we conclude that this galaxy may be a LyC leaker candidate (Fig. \ref{fig:r23_discussion}).  \\

\item For galaxies with a significant \oiiia\ detection, we derived the \oiiia/\oiiil\ ratio and the electron temperature and find that these values are similar to or larger than those predicted by nebular models with extremely low metallicity, high ionisation parameters, and constant \SFR\ at $t = 3 \times 10^8$ years. From this we conclude that a part of the extreme \ott\ emitters may have light-weighted ages of $t < 3 \times 10^8$ years (Fig. \ref{fig:te}).   \\
\end{itemize}
\label{s_conclusions}

\begin{acknowledgements}
We thank the referee for a constructive report that helped improve the paper. AV is supported by a Marie Heim-V\"{o}gtlin fellowship of the Swiss National Foundation. JB acknowledges support from the Funda\c{c}\~{a}o para a Ci\^{e}ncia e a Technologia (FCT) through national funds (UID/FIS/04434/2013) and Investigador FCT contract IF/01654/2014/CP1215/CT0003., and by FEDER through COMPETE2020 (POCI-01-0145-FEDER-007672). TC acknowledges support from the ANR FOGHAR (ANR-13-BS05-0010-02), the OCEVU Labex (ANR-11- LABX-0060), and the A*MIDEX project (ANR-11-IDEX-0001-02) funded by the "Investissements d'avenir" French government programme. JS and SM acknowledge support from The Netherlands Organisation for Scientific Research (NWO), VICI grant 639.043.409. SC gratefully acknowledges support from Swiss National Science Foundation grant PP00P2$\_$163824.
\end{acknowledgements}

\bibliographystyle{aa}
\bibliography{refs}

\appendix
\section{Distribution of galaxy properties}
\begin{figure*}
 \centering
   \includegraphics[width=\textwidth]{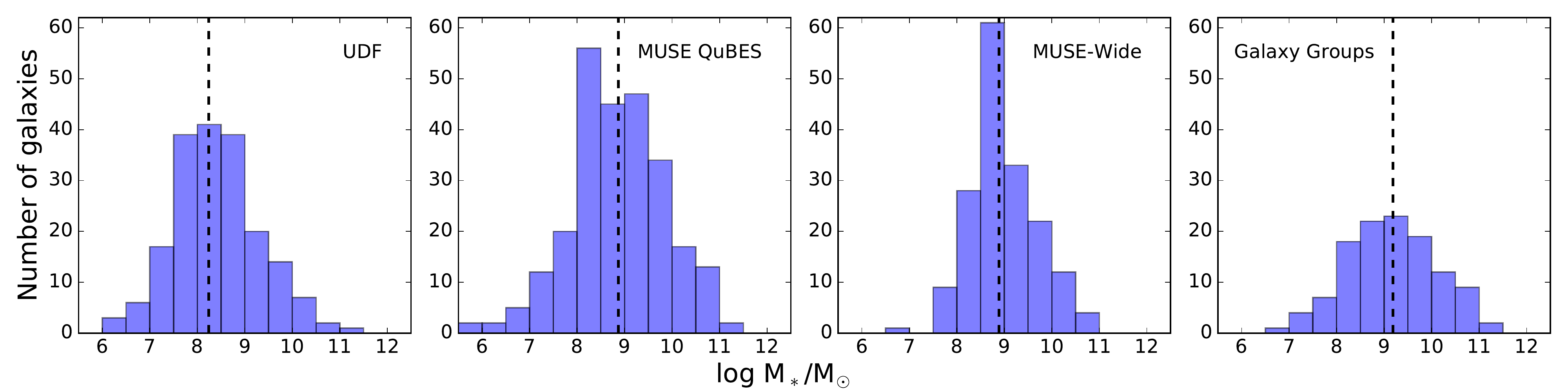}
      \caption{The distribution of the logarithm of the stellar mass, derived by standard spectral energy distribution (SED) fitting using the FAST algorithm \citep{2009ApJ...700..221K}. The median stellar mass of each survey is shown by the dashed black lines.} 
      \label{fig:app_distr_m}
\end{figure*}

\begin{figure*}
 \centering
   \includegraphics[width=\textwidth]{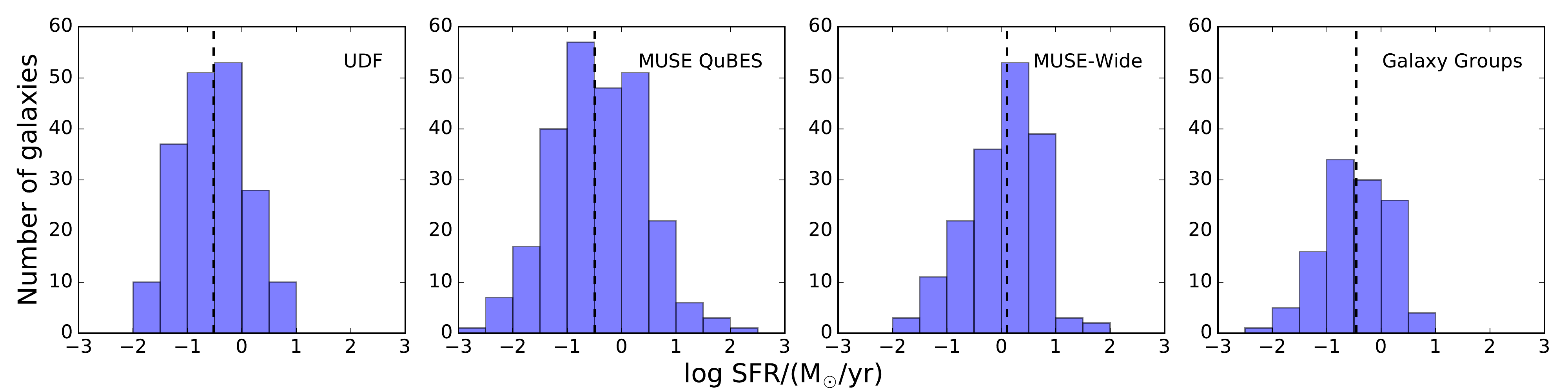}
      \caption{Distribution of the $\log$ \SFR\ estimates and the median values (dashed black lines), derived by the method from \citet{2013MNRAS.432.2112B} of the three different surveys used for this study.  } 
      \label{fig:app_distr_sfr}
\end{figure*}

In addition to the histograms of the distances to the SFMS of the four different surveys (Fig. \ref{fig:distr_ssfr}), we show here the \Mstar\ (Fig. \ref{fig:app_distr_m}) and the \SFR\ (Fig. \ref{fig:app_distr_sfr}) histograms. Because deep HST photometry is used a priori for the detection of galaxies in the UDF survey, the median \Mstar\ of the UDF is at lower stellar mass than that of the other surveys. However, the median stellar mass of the Galaxy Groups survey is higher, at $\log$ \Mstar/\Msun\ $\approx 9.2$, because this survey is aimed to observe galaxy groups that contain higher mass galaxies. The \SFR\ distributions reflect the differences of the depth of the MUSE data of the different surveys. The median \SFR\ of the shallowest survey, MUSE-Wide, is therefore approximately one order of magnitude larger than that of the other surveys.

\begin{figure*}
 \centering
   \includegraphics[width=\textwidth]{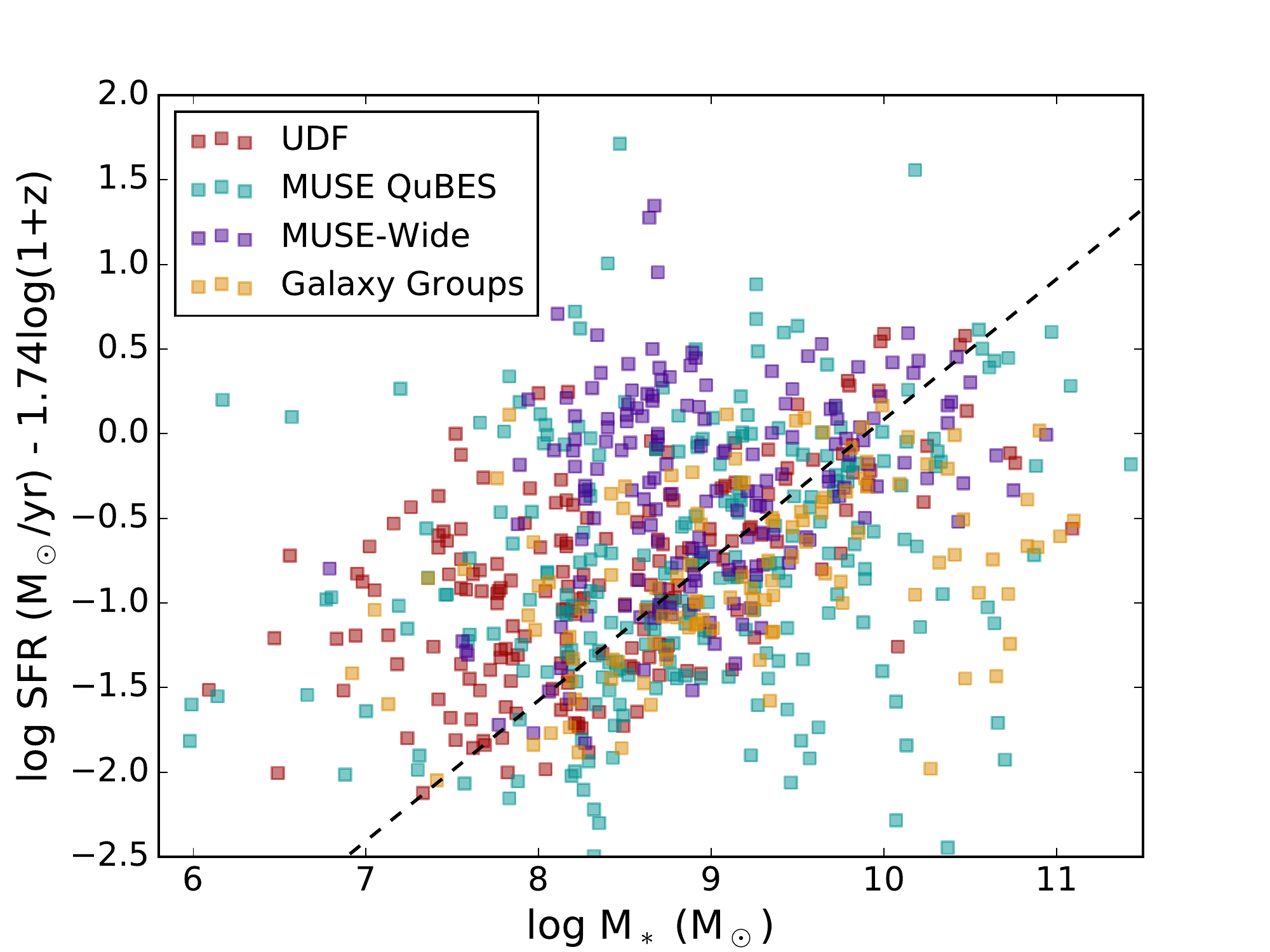}
      \caption{ \SFR\ - \Mstar\ diagram, where the \SFR\ is corrected to $z=0$ using the redshift evolution from Eq. \ref{SFMS}, similar to Fig. \ref{fig:main_sequence}. The symbols are coloured according to the survey from which the galaxies originate. }
      \label{fig:app_SFMS}
\end{figure*}

In Fig. \ref{fig:app_SFMS} we show originating survey of individual galaxies in a redshift-corrected \SFR\ - \Mstar\ diagram. Below this relation, galaxies from the MUSE QuBES and the MUSE-Wide surveys are more densely populated, which means that most of the galaxies that are not selected for the final sample are from these surveys. 

\label{app:distr}

\section{Stellar mass and flux completeness in SDSS data}
\label{app:sdss}

\begin{figure*}
 \centering
   \includegraphics[width=\textwidth]{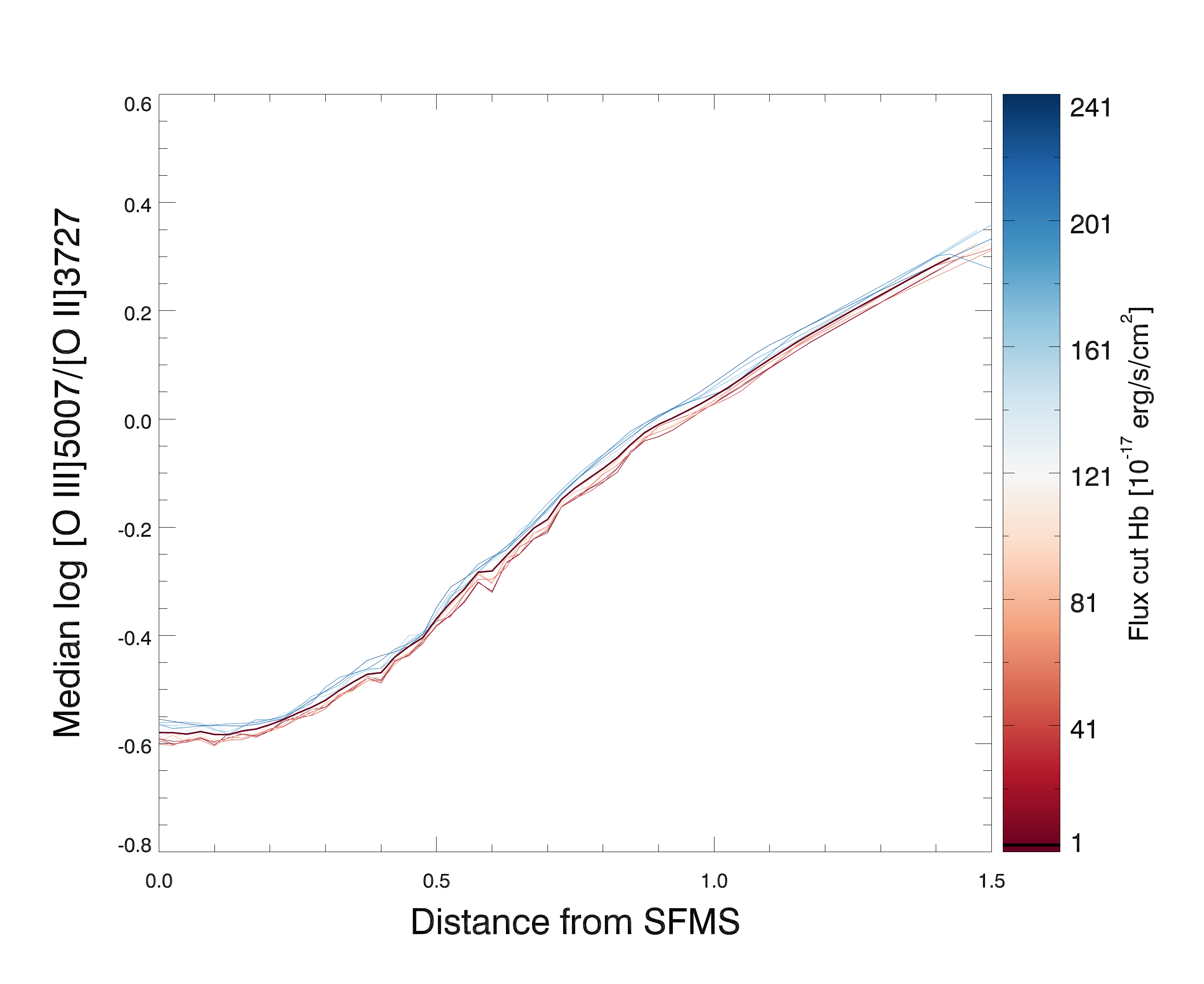}
      \caption{Median of the logarithm of \ott\ versus the distance to the SFMS of Eq. \ref{SFMS}, shifted by 0.2 in the $\log$ \SFR\ - 2.93 $\log$(1+z) direction to correct for the different flux measurements method for the SDSS data. The colours indicate the \hb\ flux cut.} 
      \label{fig:o32means}
\end{figure*}

\begin{figure*}
 \centering
   \includegraphics[width=\textwidth]{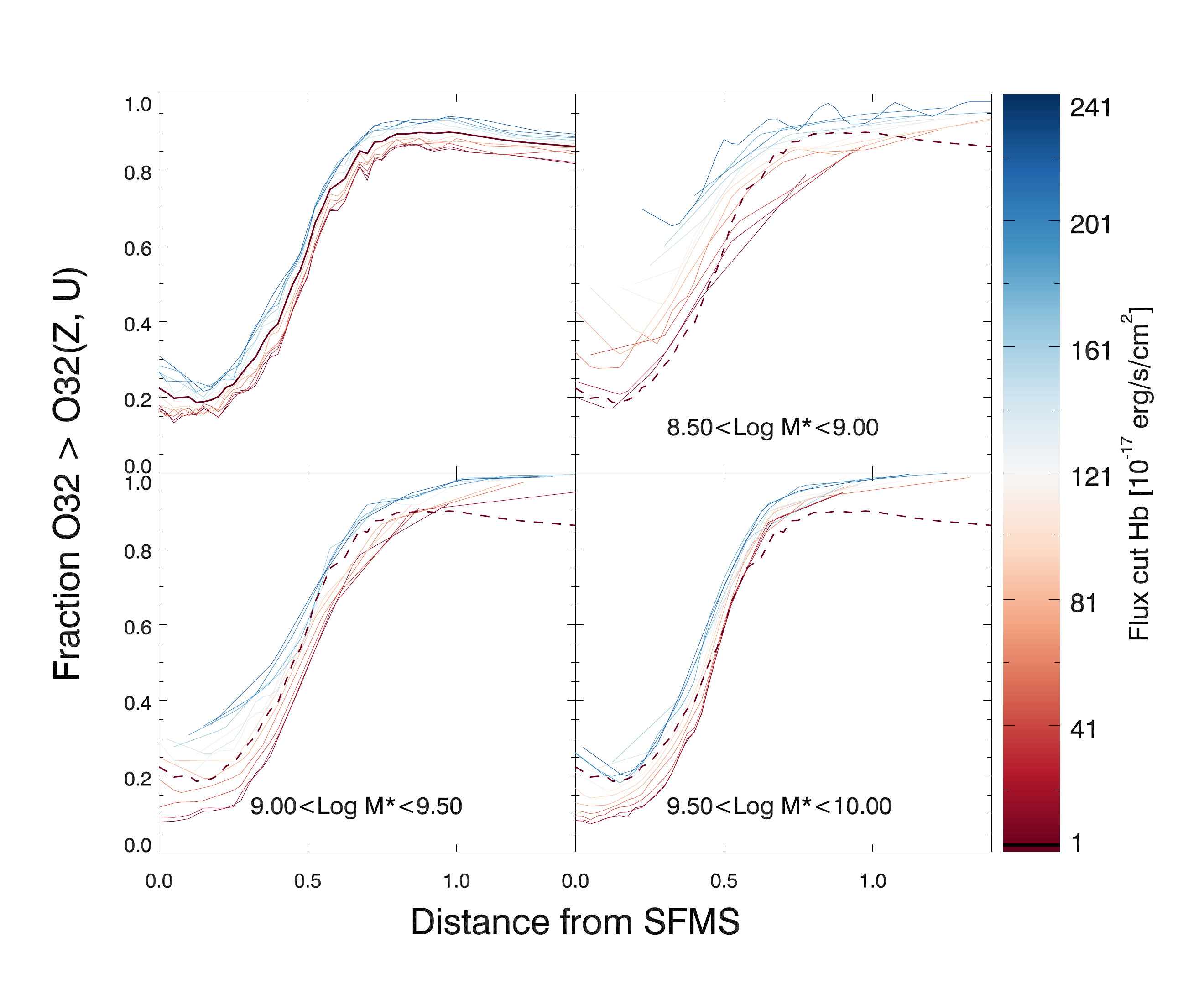}
      \caption{ Fraction of \ott\ above the metallicity-dependent threshold for high \ott\ emitters versus the distance to the SFMS. The colours are used in the same way as in the previous figure. The upper panel shows the entire SDSS sample, the other panels for a \Mstar\ range are specified in green. The median of the fraction of the entire sample from the upper left panel is presented by the black dashed line.} 
      \label{fig:o32fraction}
\end{figure*}

Here we discuss the influence of flux incompleteness and \Mstar\ incompleteness on the \ott\ ratio and the high \ott\ fractions. To study how potential incompletenesses affect our results, we use star-forming galaxies selected from the SDSS DR7 sample, which is introduced in Sect. \ref{section:sdsscomp}. We assume this sample is complete down to  $\log$ \Mstar/\Msun\ = 9 (see Fig. 4 in \citealt{2008MNRAS.388..945B}). In Fig. \ref{fig:o32means} we show the relation between the distance to the SFMS and log \ott , where we applied different flux cuts on \hb\ (indicated by the colours). This shows that the median \ott\ is independent of the flux cut and as such the \ott\ fractions from Sect. \ref{incidenceratemass} are not affected by the flux incompleteness. We demonstrate the effect of the completeness in \Mstar\ in Fig. \ref{fig:o32fraction}. Here we show the fraction of galaxies with \ott\ above the metallicity-dependent threshold, defined in Sect. \ref{sec:metalthreshold}, versus the distance to the SFMS. The top left panel shows the relations for all galaxies and in the other three panels we show the relations in different mass intervals. The upper right panel with $8.5 < \log \Mstar/\Msun < 9.0 $ is mass-incomplete, however, if we compare the two lowest panels, we see that the trends are comparable to the median of the entire sample (dashed black line) and independent of the \hb\ flux cut. This suggest that the trend that we derived between the fraction of high oxygen emitters and mass is not very sensitive to a possible mass incompleteness. 

\end{document}